\documentclass[twocolumn]{aastex631}

\usepackage{amsmath}
\usepackage{multirow}

\begin{document}

\title{No longer impossible: the self-lensing binary KIC 8145411 is a triple}

\author[0000-0001-6970-1014]{Natsuko Yamaguchi}
\affiliation{Department of Astronomy, California Institute of Technology, 1200 E. California Blvd, Pasadena, CA, 91125, USA}

\author[0000-0002-6871-1752]{Kareem El-Badry}
\affiliation{Department of Astronomy, California Institute of Technology, 1200 E. California Blvd, Pasadena, CA, 91125, USA}

\author[0000-0002-5741-3047]{David R. Ciardi}
\affiliation{NASA Exoplanet Science Institute, IPAC, California Institute of Technology, Pasadena, CA 91125, USA}

\author[0000-0001-9911-7388]{David W. Latham}
\affiliation{Center for Astrophysics ${\rm \mid}$ Harvard {\rm \&} Smithsonian, 60 Garden Street, Cambridge, MA 02138, USA}

\author[0000-0003-1298-9699]{Kento Masuda}
\affiliation{Department of Earth and Space Science, Graduate School of Science, Osaka University, 1-1 machikaneyama, Toyonaka, Osaka 560-0043, Japan }

\author[0000-0001-6637-5401]{Allyson Bieryla}
\affiliation{Center for Astrophysics ${\rm \mid}$ Harvard {\rm \&} Smithsonian, 60 Garden Street, Cambridge, MA 02138, USA}

\author[0000-0002-2361-5812]{Catherine A. Clark}
\affiliation{NASA Exoplanet Science Institute, IPAC, California Institute of Technology, Pasadena, CA 91125, USA}
\affiliation{Jet Propulsion Laboratory, California Institute of Technology, 4800 Oak Grove Drive, Pasadena, CA 91109, USA}

\author[0000-0003-4255-3650]{Samuel S. Condon}
\affiliation{Department of Physics, California Institute of Technology, 1200 E. California Blvd, Pasadena, CA, 91125, USA}

\begin{abstract}

Five self-lensing binaries (SLBs) have been discovered with data from the \textit{Kepler} mission. One of these systems is KIC 8145411, which was reported to host an extremely low mass (ELM; $0.2\,M_{\odot}$) white dwarf (WD) in a 456-day orbit with a solar-type companion. The system has been dubbed ``impossible'', because evolutionary models predict that $\sim 0.2\,M_{\odot}$ WDs should only be found in tight orbits ($P_{\rm orb} \lesssim$ days). In this work, we show that KIC 8145411 is in fact a hierarchical triple system: it contains a WD orbiting a solar-type star, with another solar-type star $\sim 700\,$AU away. The wide companion was unresolved in the Kepler light curves, was just barely resolved in Gaia DR3, and is resolved beyond any doubt by high-resolution imaging. We show that the presence of this tertiary confounded previous mass measurements of the WD for two reason: it dilutes the amplitude of the self-lensing pulses, and it reduces the apparent radial velocity (RV) variability amplitude of the WD's companion due to line blending. By jointly fitting the system's light curves, RVs, and multi-band photometry using a model with two luminous stars, we obtain a revised WD mass of $(0.53 \pm 0.01)\,M_{\odot}$. Both luminous stars are near the end of their main-sequence evolution. The WD is thus not an ELM WD, and the system does not suffer the previously proposed challenges to its formation history. Similar to the other SLBs and the population of astrometric WD binaries recently identified from Gaia data, KIC 8145411 has parameters in tension with standard expectations for formation through both stable and unstable mass transfer. The system's properties are likely best understood as a result of unstable mass transfer from an AGB star donor. 
\end{abstract}


\section{Introduction} \label{sec:intro}

Self-lensing binaries (SLBs) are eclipsing binaries which contain a compact object that gravitationally lenses the light from its companion. The resulting amplification dominates over the usual dimming by the eclipse, resulting in an overall brightening when the compact object transits in front of the star. Using data from the \textit{Kepler} mission, five self-lensing binaries have been  discovered. The first of these was KOI-3278, discovered by \citet{Kruse2014Sci}. Another three were discovered by \citet{Kawahara2018AJ}, who also identified a fourth candidate which was later confirmed by \citet{Masuda2019ApJL} through radial velocity (RV) follow-up.

\begin{figure*}
    \centering
    \includegraphics[width=0.97\textwidth]{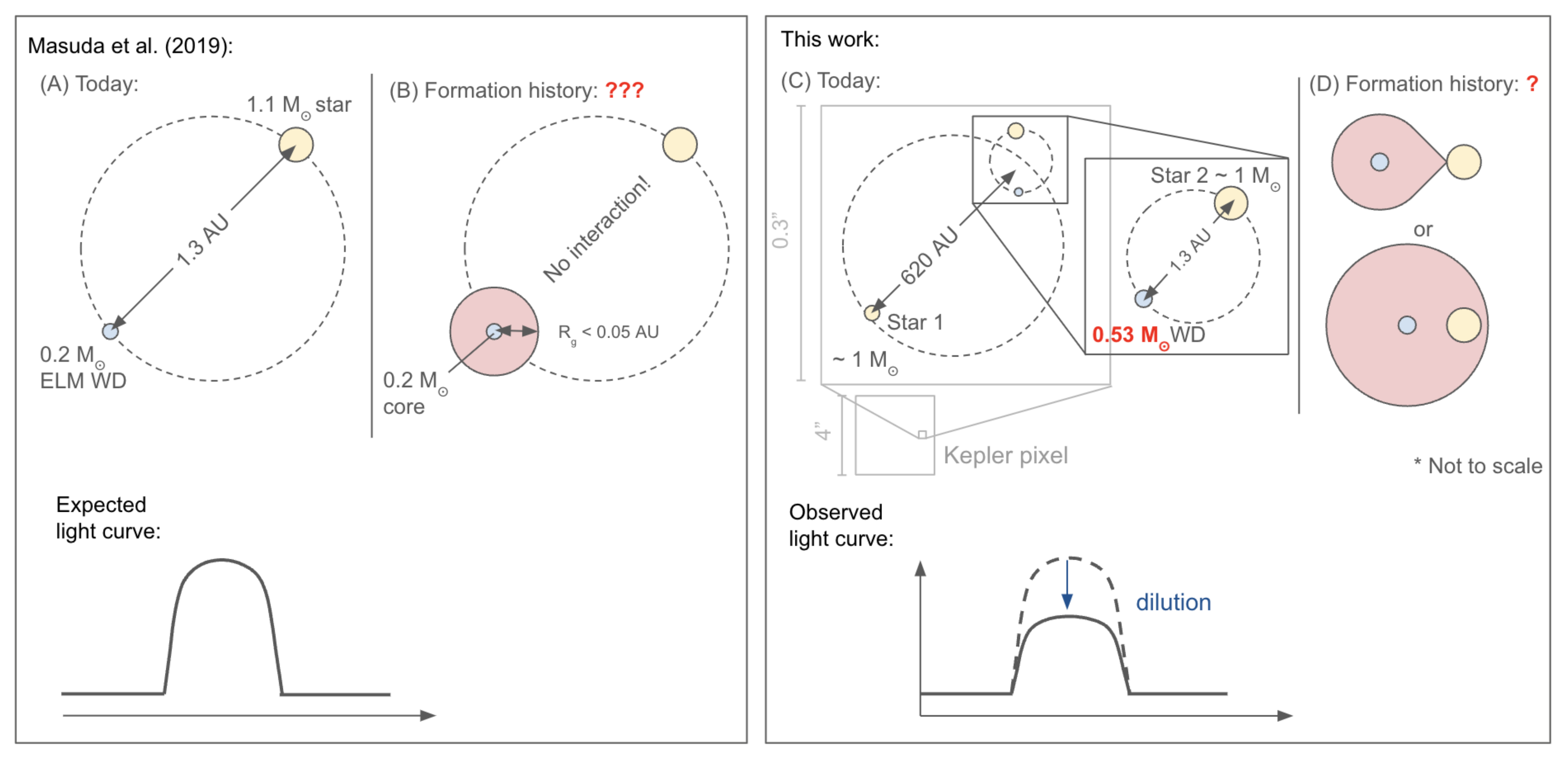}
    \caption{Schematic drawings of KIC 8145411 based on the original analysis of the system as a binary with a single luminous star by \citet{Masuda2019ApJL} (\emph{left}) and the updated results from this work for a hierarchical triple with two luminous stars (\emph{right}). On the left of each panel (A, C), we show the present-day system, and on the right (B, D), we illustrate the difficulties with explaining its formation (see text). For reference, we show that the size of the Kepler pixel is larger than the orbit of the tertiary and thus the two stars are not resolved. On the bottom, we show the effect that this dilution has on the observed pulse.}
    \label{fig:illustration}
\end{figure*}

These systems span a range of orbital periods from a few months to years, which are not predicted by standard binary population synthesis models for white dwarfs (WDs) of the observed masses. Having the shortest orbital period of 88 days, KOI-3278 is generally thought to have formed through common envelope evolution (CEE). However, its period is still much longer than the majority of other known post-common envelope binaries (PCEBs), requiring minimal orbital shrinkage to have taken place during CEE, and thus highly efficient envelope ejection. This has led to much work exploring the energetics of CEE to try to explain its orbit \citep[e.g.][]{Zorotovic2014A&A, Belloni2024arXiv}. Meanwhile, at least two out of the three systems of \citet{Kawahara2018AJ} have orbital periods that are shorter than expected for stable mass transfer  (MT) systems, while being significantly longer than traditional PCEBs, making their formation histories also elusive. But perhaps the most puzzling SLB is the final system, KIC 8145411. \citet{Masuda2019ApJL} found that it hosts an extremely low-mass (ELM) WD with a mass of $0.2\,M_{\odot}$ in a $456$-day orbit around a $1.1 \, M_{\odot}$ solar-type star. Its orbit is much wider than the majority of known ELM WD binaries (with WD masses $\lesssim 0.3\,M_{\odot}$ and periods $\lesssim 1$ day; \citealt{Brown2020ApJ}). This makes it difficult to explain its formation, as mass transfer from the WD progenitor must have begun early on the RGB to form a WD of just $0.2\,M_{\odot}$, but such a star would not have been large enough for interaction to occur given the large separation. A schematic diagram of this scenario is shown on the left panel of Figure \ref{fig:illustration}. In view of this challenge, alternative formation histories, such as a merger of an inner binary \citep{Masuda2019ApJL} or more recently, dynamical assembly \citep{Khurana2023ApJ}, have been considered. 

SLBs are valuable because they are sensitive to a similar range of orbital periods as the large sample of main sequence (MS) + WD binaries from the recent third data release from the Gaia mission \citep{Shahaf2024MNRAS}. SLBs provide a complementary sample with a relatively simple selection function, which depends primarily on the eclipse (i.e. edge-on alignment) probability. This is in constrast to Gaia's multi-step pipeline to determine sources that are published with astrometric binary solutions \citep[see Figure 1 of][]{Halbwachs2023A&A}. This is key in the calculations of intrinsic space densities and rates. 

In this work, we present evidence that KIC 8145411 is a triple, with an inner solar-type star + WD binary and an outer solar-type companion, as illustrated on the right panel of Figure \ref{fig:illustration}. Taking into account the light from the outer companion, both reduces the semi-amplitude of the radial velocities (RVs) and dilutes the pulse (i.e. the positive self-lensing signal) from the light curve, pushing up the mass of the WD so that it is no longer an ELM WD. In Section \ref{sec:discovery}, we summarize the original discovery of KIC 8145411 and its previously determined orbital parameters by \citet{Masuda2019ApJL}. We describe the data in Section \ref{sec:data} and detail the joint fitting in Section \ref{sec:fitting}. The results of the analysis and the implications on the formation history of the system are presented in Section \ref{sec:results}. Finally, we conclude with a summary of our findings in Section \ref{sec:conclusion}. 

\section{Original discovery} \label{sec:discovery}

KIC 8145411 was first identified as a possible SLB by \citet{Kawahara2018AJ}, who searched for pulses in the long cadence PDCSAP light curves from \textit{Kepler}'s primary mission. They identified a total of three confirmed SLBs. They also identified KIC 8145411 -- which had two possible orbital solutions due to a gap in the light curve data-- as an unconfirmed candidate. The system was later confirmed through further RV follow-up in \citet{Masuda2019ApJL}, who jointly fitted the RVs and light curves assuming a binary with one luminous companion, and reported that it hosts a $0.2\,M_{\odot}$ ELM WD and a $1.1\,M_{\odot}$ primary in a 456-day orbit with a relatively low eccentricity of $\sim 0.14$. This corresponds to a semi-major axis of $\sim 1.3\,$AU. This configuration is illustrated on the left panel of Figure \ref{fig:illustration}. 

The parameters inferred by \citet{Masuda2019ApJL} are difficult to understand in the context of isolated binary evolution models: a red giant progenitor to a $0.2\,M_{\odot}$ white dwarf would have a radius of only $\sim 9\, R_{\odot}$, which is too small for mass transfer to have occurred in such a wide orbit. Alternative scenarios for the formation of ELM WDs do exist, such as the merger of an inner binary \citep{Vos2018MNRAS} but the relative circularity of this orbit is suggestive of binary interaction. 

\section{Data} \label{sec:data}

Throughout the paper, the outermost star will be referred to as star 1, as it is the brighter star in the G band. It orbits an inner binary containing star 2 and the WD. Quantities corresponding to each object will be denoted by subscripts 1, 2, and WD, respectively.

\subsection{Gaia DR3} \label{ssec:gaia}

We noticed that there are two Gaia DR3 \citep{GaiaCollaboration2016A&A, GaiaCollaboration2023A&A} sources at the coordinates of KIC 8145411, separated by $\sim 0.3$ arcseconds. The source IDs for star 1 and 2 are 2105324940517591808 and 2105324936217850624, respectively. These both have 2-parameter (i.e. position only) solutions, meaning no parallax or proper motion constraints are available. Only $\sim 0.1$\% of Gaia sources as bright as these sources have 2-parameter solutions \citep{GaiaCollaboration2021A&A}, so we suspect the 5-parameter solutions failed as a result of the close proximity of the two sources (this effect has been shown by \citealt{Tokovinin2023AJ}). There were {\it no} Gaia sources near KIC 8145411 in Gaia DR2, likely because of the difficulty of resolving the close pair. One might wonder whether Gaia could have detected the same source twice, but the high \texttt{ipd\_frac\_multi\_peak} values (a measure of the percentage of scans in which more than one peak was detected) of both Gaia sources are indicative of a marginally-resolved binary (\citealt{Lindegren2021A&A}; as do the spectra, see Section \ref{ssec:tres_spectra}). Moreover, the reported G band fluxes are likely reliable as they roughly sum to the unresolved flux of the two stars in other optical bands (Section \ref{ssec:2mass_panstarrs}). 

\subsection{Kepler}

We use the Kepler light curve for KIC 8145411, only keeping data within 4 days of the two self-lensing pulses, identified by \citet{Masuda2019ApJL}. The light curves were de-trended using a second-order polynomial. 

The secondary eclipse is not detected. This places an upper limit on the WD temperature (see Section \ref{sec:results}). 

\subsection{2MASS and Pan-STARRS photometry} \label{ssec:2mass_panstarrs}

We use 2MASS JHKs \citep{Skrutskie2006AJ} and Pan-STARRS1 \textit{griz} \citep{Chambers2016arXiv} photometry of this source. The two stars are not resolved in either of these surveys and therefore the photometry corresponds to the total flux from both stars. The combination of multi-band flux ratios as well as the total apparent magnitudes in these bands will allow us to place constraints on the location of the two stars on an isochrone, constraining the distance, which is not constrained by Gaia (see Section \ref{sec:results}). All of the photometric data used in our analysis is summarized in Table \ref{tab:photometry_summary}. 

\begin{table*}[htb!]
  \begin{tabular}{l l|l l|l l|l l}
    \hline
    \multicolumn{2}{c|}{2MASS (unresolved)} &
    \multicolumn{2}{c|}{Pan-STARRS (unresolved)} &
    \multicolumn{2}{c|}{Gaia (resolved)} &
    \multicolumn{2}{c}{PHARO (resolved)} \\
    \hline
    J & 13.443 $\pm$ 0.021 & g & 15.098 $\pm$ 0.003 & $G_1$ & 15.2208 $\pm$ 0.004 & $J_1$  & 14.052 $\pm$ 0.021 \\
    H & 13.078 $\pm$ 0.022 & r & 14.601 $\pm$ 0.002 & $G_2$ & 15.42 $\pm$ 0.01 & $J_2$  & 14.362 $\pm$ 0.022 \\
    Ks & 13.009 $\pm$ 0.023 & i & 14.463 $\pm$ 0.003  & & & $H_1$ & 13.665 $\pm$ 0.022\\
    & & z & 14.408 $\pm$ 0.003 & & & $H_2$ & 14.026 $\pm$ 0.023 \\
    & & & & & & $Ks_1$ & 13.590 $\pm$ 0.023\\
    & & & & & & $Ks_2$ & 13.965 $\pm$ 0.024 \\
    \hline
  \end{tabular}
  \caption{Summary of the photometry used in our analysis. 2MASS and Pan-STARRS values are unresolved magnitudes of the two stars combined, while for Gaia and PHARO, we report the deblended magnitudes of each resolved star. The Gaia G band magnitude errors were calculated using the reported flux errors.}
  \label{tab:photometry_summary}
\end{table*}

\subsection{Palomar Near-Infrared Adaptive Optics Imaging} \label{ssec:PHARO}

\begin{figure*}
    \centering
    \includegraphics[width = 0.8\textwidth]{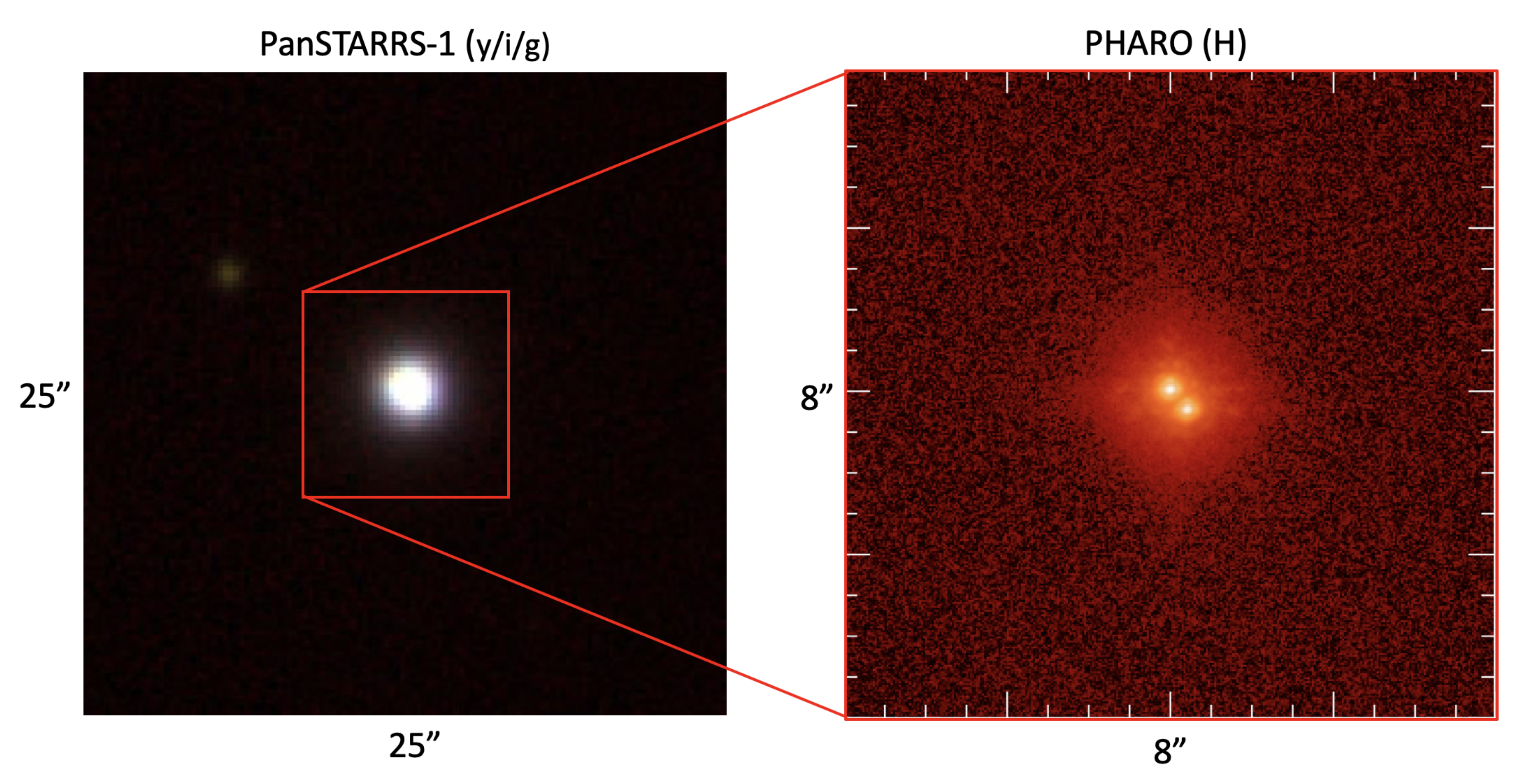}
    \caption{Image of KIC 8145411 in PanSTARRS-1 (stacked y/i/g, 25" $\times$ 25") and a zoomed-in image taken by PHARO (H band, 8" $\times$ 8"). We see that the two luminous components are resolved by PHARO, separated by $\sim 0.3$".}
    \label{fig:pharo_image}
\end{figure*}

To confirm that the system is indeed a triple, we obtained high-resolution, near-infrared adaptive optics imaging at Palomar Observatory utilizing the PHARO instrument \citep{hayward2001} behind the natural guide star adaptive optics (AO) system P3K \citep{dekany2013} on 2022-04-21 UT. The AO data were acquired in a standard 5-point quincunx dither pattern with steps of 5\arcsec\ in the J, H, and Ks filters.  Each dither position was observed three times, offset in position from each other by 0.5\arcsec\ for a total of 15 frames; with an integration time of 15 seconds per frame, respectively, for total on-source times of 225 seconds. PHARO has a pixel scale of $0.025\arcsec$ per pixel for a total field of view of $\sim25\arcsec$, and resolution of the imaging, as measured by the FWHM of the point sources, is 0.099\arcsec, 0.088\arcsec, and 0.101\arcsec, respectively for the J, H, and Ks filters. The final combined mosaic (Figure \ref{fig:pharo_image}) clearly resolves the two stars in all three bands, separated by $\sim0.32\arcsec$. The infrared flux observations provide flux ratios between the stars in 3 bands. The flux ratios for the J, H, and Ks filters are $\Delta J = 0.310\pm 0.008$, $\Delta H= 0.361 \pm 0.007$, and $\Delta Ks = 0.375\pm 0.007$ mag, respectively (Table \ref{tab:photometry_summary}). The separations and position angles of the two stars in the three bands, as well as their locations on a JHKs color-color plot can be found in Appendix \ref{appendix:pharo_extra}. 

\subsection{TRES spectra} \label{ssec:tres_spectra}

\begin{figure*}
    \centering
    \includegraphics[width=0.53\textwidth]{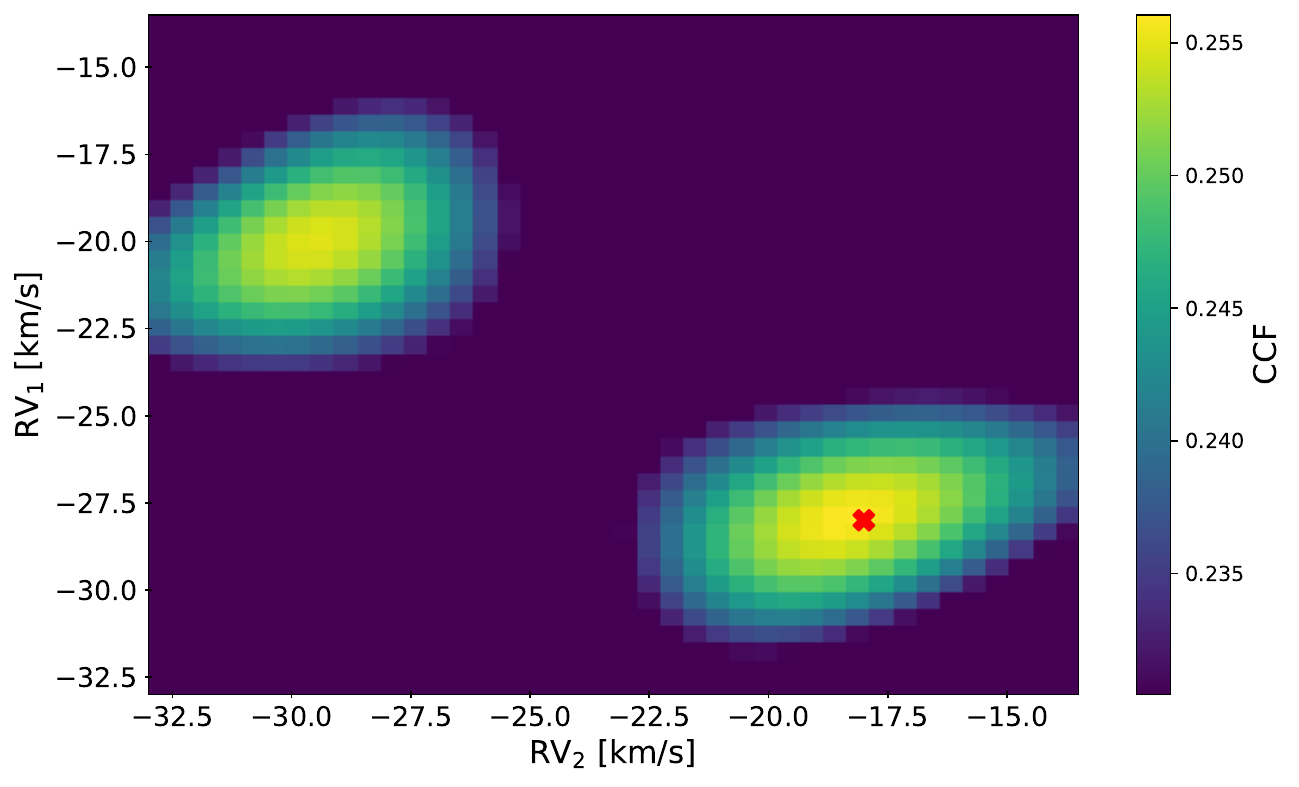}
    \includegraphics[width=0.43\textwidth]{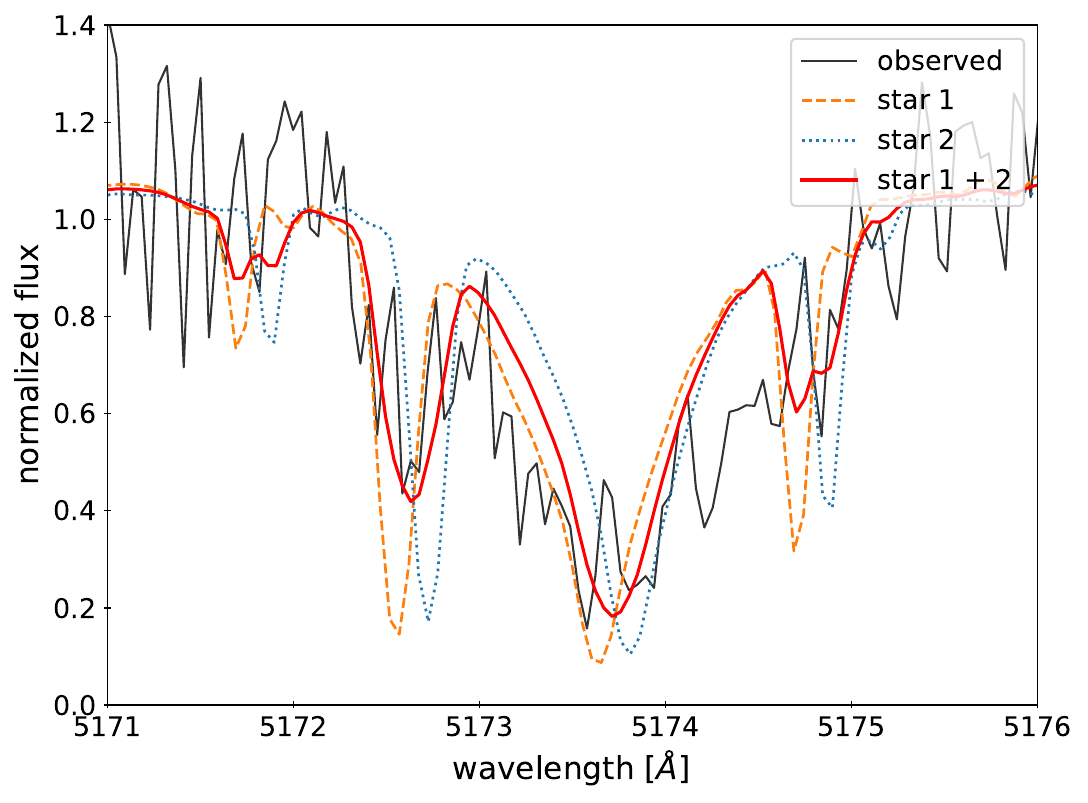}
    \caption{\emph{Left}: CCF between the observed and template spectrum for one epoch (BJD 2458210.910), as a function of the radial velocities of star 1 and 2. The location of the maximum CCF value is marked with a red cross. This is used as the starting point in the optimization. We see a second peak where the two RVs are reversed, which corresponds to switching the identities of the two stars. \emph{Right}: Zoom-in of the spectrum on one of the Mg triplet lines (rest wavelength $5172.684\,$\AA). The dashed and dotted lines are the template spectra for star 1 and star 2 respectively, each shifted by their radial velocities. The red solid line is the combined template, which is the sum of the two single star spectra, scaled by their flux ratios and normalized.}
    \label{fig:ccf}
\end{figure*}

\begin{figure*}
    \centering
    \includegraphics[width=0.9\textwidth]{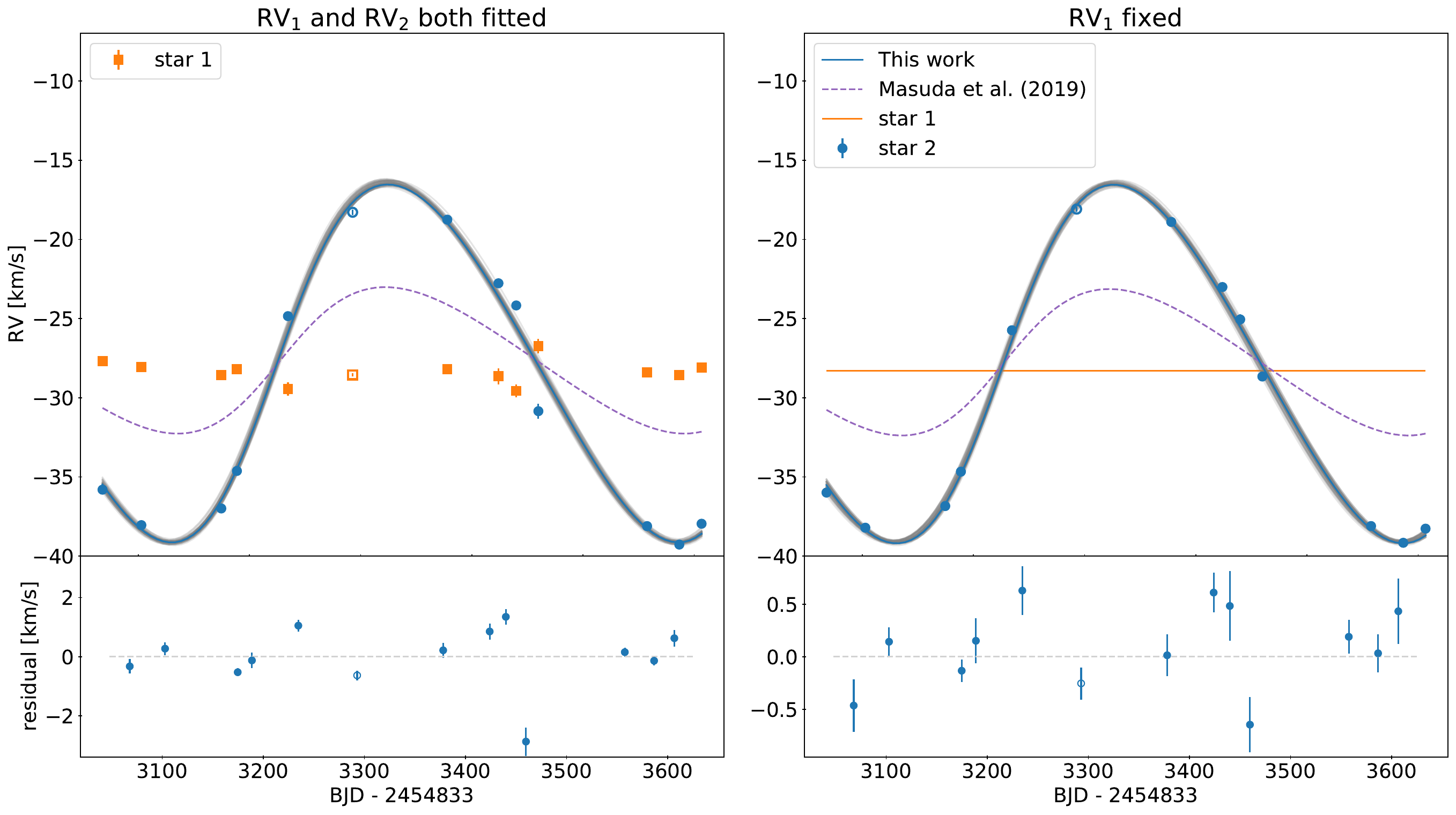}
    \caption{Measured RVs using two methods  (Section \ref{ssec:tres_spectra}). Note that the RV point taken at BJD$_{\rm TDB} - 2454833 = 5571.9343$ (unfilled marker, located at x-coordinate $\sim 3293$) has been shifted back 5 periods for easier visualization. \emph{Left}: fitting RVs of both star 1 and star 2; \emph{Right}: keeping the RVs of star 1 fixed at a constant value and only fitting those of star 2. We also plot the best-fit models from our joint fitting (blue solid line), along with 50 random models from the posterior (Section \ref{sec:results}). The lower panels plot the RV residuals from the best-fit models. For reference, we also plot the solution from the joint RV and light curve fitting from \citet{Masuda2019ApJL} assuming a single luminous star (with a constant RV offset of our best-fit $\gamma$). We see that that the semi-amplitudes of our measured RVs are significantly larger, which lead to higher inferred WD masses.}
    \label{fig:rv_models}
\end{figure*}

To measure RVs, we use 12 spectra already presented by \citet{Masuda2019ApJL} and one new spectrum, which were all obtained with the Tillinghast Reflector Echelle Spectrograph \citep[TRES;][]{Furesz2008}, mounted on the 1.5\,m Tillinghast Reflector telescope at the Fred Lawrence Whipple Observatory (FLWO) in Arizona. As described in \citet{Masuda2019ApJL}, the exposure times were $\sim 1$ hour, and the resulting spectra have resolution $R \sim 44,000$ and signal-to-noise $\sim 15-20$. Since the 0.3" angular separation of the two stars is much smaller than the 2.3" TRES fiber diameter, we expect both stars to contribute to the spectra according to their flux ratio.   

We obtain Kurucz synthetic spectra (from the \texttt{BOSZ} grid \citealt{Bohlin2017}) for star 1 and star 2 with a range of effective temperatures $T_{\rm eff, 1}$ and $T_{\rm eff, 2}$ with $R = 50,000$ (comparable to that of TRES). We used templates with log(g) = 4.0 and solar abundances. These were each shifted over a coarse grid of RVs (intervals of $0.5\,\mbox{km s}^{-1}$, starting with a range from $\sim -10$ to $-50 \mbox{km s}^{-1}$ based on RVs from \citealt{Masuda2019ApJL}), scaled by their relative flux contributions in the Gaia G band, and summed to predict the combined spectrum. This combined spectrum was then cross-correlated with the observed TRES spectrum, and the best RVs from the grid were taken as the starting point in the 2D optimization (we used the Nelder-Mead algorithm in scipy's optimize.minimize function) to measure the final best-fit pair of RVs. We neglect the contribution of light from the WD based on the lack of detection of the secondary eclipse in the light curve. On the left panel of Figure \ref{fig:ccf}, we plot the cross-correlation function (CCF) between the observed and template spectra against the RVs of star 1 and star 2 for one epoch. On the right, we show the individual template spectra of the two stars and their combined spectrum for the same epoch. Note that the spectral features of the combined spectrum are broadened by the velocity offset between the two stars, which is likely responsible for the large value of $v\mbox{sin}i$ measured by \citet{Masuda2019ApJL}. 

We measured RVs for orders 15 - 30, which span a wavelength range of $\sim 4640$ to $5960\,$\AA, over which the flux ratios are not expected to depart much from that in the Gaia G band ($F_1/F_2 \sim 1.2$, see below). This is consistent with seeing no visible trends in the RVs across the orders (Figure \ref{fig:rvs_across_orders}, Appendix \ref{appendix:rvs_across_orders}). The final best-fit SEDs of the stars (Section \ref{sec:results}, Figure \ref{fig:seds}) also provide flux ratios of $\sim 1.1 - 1.2$ across this wavelength range. We take the best-fit RVs as the median across these orders and the errors as the standard deviation over the square root of the number of orders. We remove anomalous orders for which the RV deviates by more than 15\% from the median at each epoch. 

Initially, we let RVs vary for both star 1 and star 2. As plotted on the left panel of Figure \ref{fig:rv_models}, we find that one star -- star 1 -- has roughly constant RVs across epochs (orange square markers), at a median value of $\sim -28.3 \mbox{km s}^{-1}$. This star is presumably the outer tertiary, which should have an orbital period of order 10,000 years (given the angular separation and distance estimate from the SED), so its RV is expected be constant over the time spanned by our observations. This is also the star that contributes more light in our fitting. However, there is some degeneracy between a pair of RVs and the reversed pair (as seen by the two peaks in the CCF on the left panel of Figure \ref{fig:ccf}). This effect is reduced by using a narrower range of RVs for star 1, but at times where the RV curves cross, this degeneracy can occasionally result in the incorrect assignment being made. In practice, the majority of these points are not used as they are removed by the 15\% outlier cut (points that are removed are marked as crosses in Figure \ref{fig:rvs_across_orders} of Appendix \ref{appendix:rvs_across_orders}). As a test, we also perform the same calculations without the outlier removal and find that most parameters agree to within $<2\sigma$ (Appendix \ref{appendix:tests}). Since the RV of star 1 is roughly constant, we then fit the spectra by fixing it to $-28.3 \mbox{km s}^{-1}$ and only varying that of star 2. The resulting RVs are plotted on the right panel of Figure \ref{fig:rv_models}, and the median values and errors are tabulated in Table \ref{tab:rvs}. Unless otherwise stated, we will use this set of RVs throughout the rest of the paper (though we also report results using the RVs from the fitting of both stars in Appendix \ref{appendix:tests}). We note that this choice has little effect on the semi-amplitude and therefore on the inferred WD mass, but we obtain smaller RV uncertainties and a better Keplerian fit for star 2 when we fix the RVs of star 1 (see residuals in Figure \ref{fig:rv_models}). 

\begin{figure}
    \centering
    \includegraphics[width = 0.93\columnwidth]{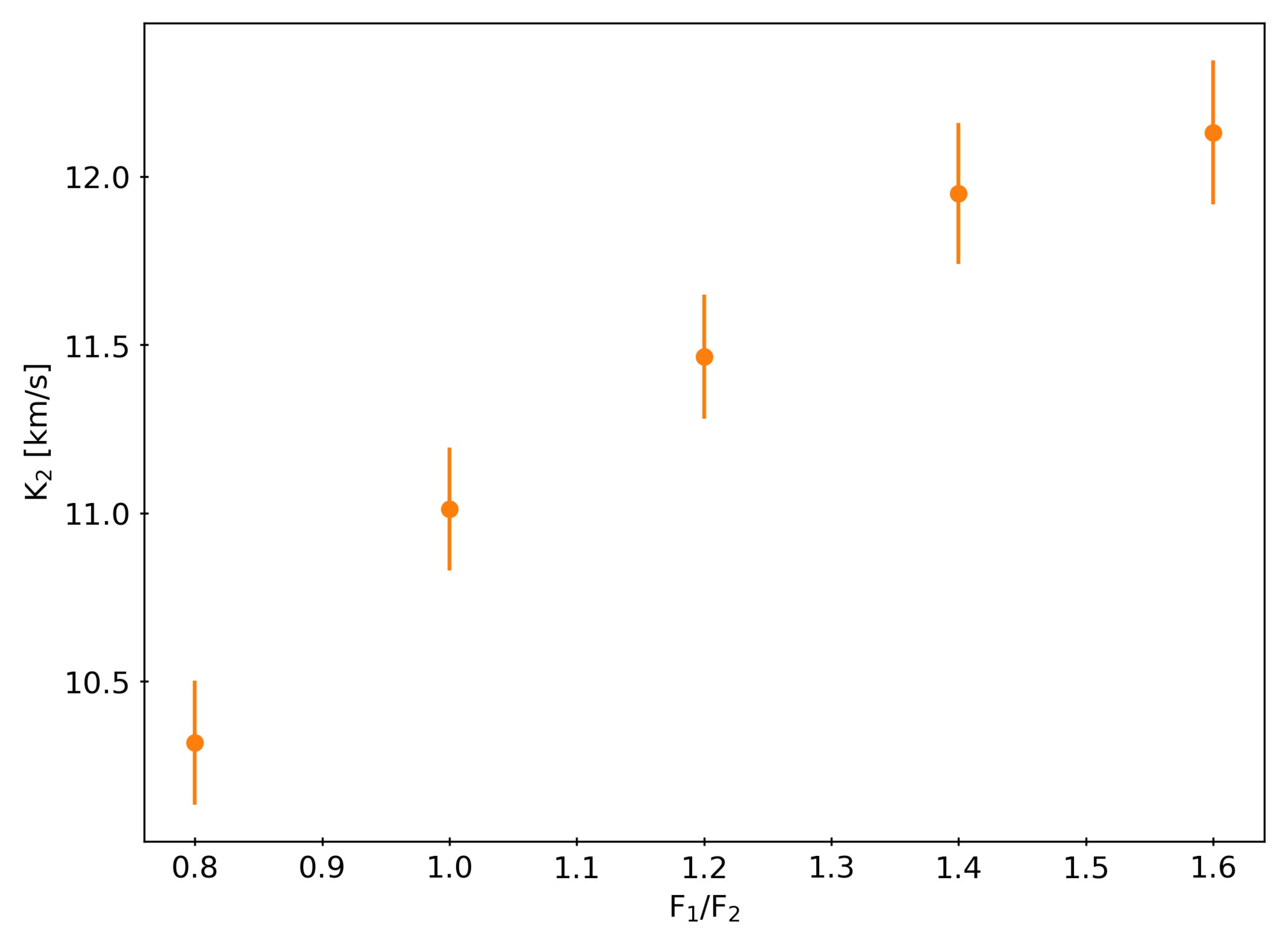}
    \caption{The best-fit semi-amplitudes of the measured RVs of star 2 against the assumed flux ratio between the two stars. The flux ratio obtained from the Gaia G band photometry is $\sim 1.2$. The range in semi-amplitudes shown here corresponds to an uncertainty in the WD mass of $\sim \pm 0.06 M_{\odot}$, while a conservative uncertainty of 10\% on the flux ratio translates to a $\sim \pm 0.02 M_{\odot}$ uncertainty in the mass.}
    \label{fig:Kstar}
\end{figure}

We obtain the flux ratio, $F_1/F_2$, between the two stars from their Gaia G band magnitudes, which gives us a value of $\sim 1.2$. While we believe that this is reasonably reliable (Section \ref{ssec:gaia}), we test the effect of varying the flux ratio on the measured RVs. As expected, if star 1 is contributing more of the light, then the true motion of star 2 is larger. Figure \ref{fig:Kstar} plots the RV semi-amplitude of star 2 (obtained from fitting just the RVs; Section \ref{ssec:rv_model}) against the assumed flux ratio. We plot a large range of flux ratios from 0.8 to 1.6 over which the semi-amplitude varies by $\sim 2 \mbox{km s}^{-1}$. For an error in the flux ratio of $\sim 0.1$ (large, given the small uncertainties in the G band magnitudes), the corresponding systematic error in the WD mass is $\lesssim 0.02 M_{\odot}$. This is comparable to the statistical error resulting from the joint fitting (Section \ref{sec:fitting}) reported in Table \ref{tab:bestfit_params}, which we quote in the rest of this paper. 

We also explored the effects of varying the effective temperatures, $T_{\rm eff, 1}$ and $T_{\rm eff, 2}$, of the two stars. We varied these over a grid ranging from 5000K to 6000K in steps of 250K (motivated by the temperature obtained from the spectral analysis from \citealt{Masuda2019ApJL}). We find that the fits to the spectra were slightly better if $T_{\rm eff, 2} > T_{\rm eff, 1}$, but that the temperature dependence of the RV curve, and in particular the semi-amplitude (which primarily determines the WD mass), is secondary to the effect of the flux ratio. Thus, in the following analysis, we take RVs measured assuming $T_{\rm eff, 1} = 5500\,$K and $T_{\rm eff, 2} = 5750\,$K. This is also roughly consistent with the values that result from the joint fitting described in Section \ref{sec:fitting} (and reported in Table \ref{tab:stars_summary}). 

\begin{table}
    \centering
    \begin{tabular}{c c}
         \hline
          BJD$_{\rm TDB}$ - 2454833 & RV$_2$ [km/s]  \\
         \hline
         3067.8163 & -35.99 $\pm$ 0.25 \\
         3102.7569 & -38.20 $\pm$ 0.14 \\
         3174.6401 & -36.83 $\pm$ 0.11 \\
         3188.7021 & -34.66 $\pm$ 0.21 \\
         3234.6041 & -25.74 $\pm$ 0.23 \\
         3377.9103 & -18.90 $\pm$ 0.20 \\
         3423.9274 & -23.01 $\pm$ 0.19 \\
         3439.9133 & -25.05 $\pm$ 0.33 \\
         3459.8197 & -28.66 $\pm$ 0.26 \\
         3557.6000 & -38.10 $\pm$ 0.16 \\
         3586.5810 & -39.15 $\pm$ 0.18 \\
         3606.6118 & -38.26 $\pm$ 0.31 \\
         5571.9343 & -18.09 $\pm$ 0.15 \\
         \hline
    \end{tabular}
    \caption{Best-fit RVs of star 2 across epochs. Here, we have fixed the RV of star 1 at $-28.3$ km/s.}
    \label{tab:rvs}
\end{table}

\section{Joint fitting} \label{sec:fitting}

\subsection{Pulse model} \label{ssec:pulse_model}

We model the pulse as described in \citet{Kawahara2018AJ}. It is made up of two components: the amplification due to self-lensing and the dimming due to eclipse. The self-lensing signal is modeled by an inverted eclipse with the depth scaled by a factor of two and using the Einstein radius of the WD as the radius of the eclipsing object \citep{Agol2003ApJ}. Following \citet{Masuda2019ApJL}, we used the \texttt{pytransit} python package \citep{Parviainen2015MNRAS} to model both signals.

We account for dilution of the pulse by the presence of star 1 using the flux ratio $\mathcal{S} = F_1/F_2$:
\begin{equation}
    f_{\rm model} = \frac{f_{\rm 2} + \mathcal{S}}{1 + \mathcal{S}}
\end{equation}
where $f_{\rm 2}$ is the flux from star 2 alone. If dilution is not accounted for, the height of the pulse is underestimated, which leads to an underestimated semi-major axis and WD mass, all else being equal. Here, we take the flux ratio in the Gaia G band, which peaks at $\sim 6000\,$\AA, similar to the Kepler passband (this is also consistent with the ratio we get from the best-fit SEDs; Figure \ref{fig:seds}). 

Unlike \citet{Masuda2019ApJL}, we do not fit for the quadratic limb-darkening coefficients of star 2 but instead use values for $T_{\rm eff} = 5750\,\mbox{K}$, $\mbox{log(g)} = 4.0$, $Z = 0$, and $\xi = 0.0$ from the \citet{Claret2011A&A} table. We choose to do this as these coefficients are not well constrained by the light curve and the results are relatively insensitive to small changes in the assumed values (but we report the effect of taking values for $T_{\rm eff} = 5500\,\mbox{K}$ in Table \ref{tab:bestfit_params_tests_2}). Following \citet{Kawahara2018AJ}, we do not fit for the WD radius independently, instead using the mass-radius relation \citep{Nauenberg1972ApJ}. The two pulses were phase-folded to be fitted with a single mid-eclipse time $t_{\rm 0}$ and orbital period but were fit with separate normalizations, $c_1$ and $c_2$. The duration of the pulse places a joint constraint on the orbital inclination and radius of the WD companion -- it is longer if the system is more edge-on or the star being transited is larger (see Figure \ref{fig:pulse_duration}). In this case, the duration requires that the radius of the star be at least $\sim 1.3\,R_{\odot}$. 

The log-likelihood function is given by:
\begin{equation} \label{eqn:logL_pulse}
    \ln \mathcal{L}_{\rm pulse} = - \frac{1}{2}  \sum_i \frac{\left({f}_{\rm model} - {f}_{\mathrm{obs}, i}\right)^2}{\sigma^2_{f, i}}
\end{equation}
where ${f}_{\mathrm{obs}, i}$ and $\sigma_{f, i}$ are the normalized flux and the corresponding error from the light curve. 

\subsection{SED model} \label{ssec:sed_model}

We fit the SED using \texttt{MINEsweeper} \citep{Cargile2020ApJ} which is a code designed for the joint modeling of photometry and spectra. Here, we only use its photometric modeling capabilities. Given a mass, metallicity, equivalent evolutionary phase (EEP), and distance $d$, it returns photometry in a range of filters and various other stellar parameters, including the radius and effective temperature. For the extinction, we assume a \citet{Cardelli1989ApJ} extinction law with $R_V = 3.1$. We use the Bayestar2019 3D dust map \citep{Green2019ApJ} which provides $E(g-r)$ (approximately equal to $E(B-V)$; \citealt{Schlafly2011ApJ}) as a function of distance. We assume solar metallicty. 

EEP is a monotonic function of age and it allows evolutionary tracks to be efficiently sampled when constructing isochrones \citep[for details, see ][]{Dotter2016ApJS}. However, two stars that are born as a binary are expected to share the same age, not necessarily the same evolutionary state. Therefore, we fit EEPs for star 1 and 2 separately and enforce that $|$log(age$_1$) - log(age$_2$)$|$ $< 0.02$. We show the result of removing this assumption in Table \ref{tab:bestfit_params_tests_2}. 

We predict the 2MASS JHKs (we take the PHARO JHKs to be the same), Pan-STARRS griz, and Gaia G band apparent magnitudes for each star. The combined 2MASS and Pan-STARRS photometry are compared to the observed (unresolved) values. We also fit the flux ratios between the two stars from the PHARO observations and Gaia G band. In other words, the log-likelihood function is:
\begin{equation} 
\begin{split}
 \ln \mathcal{L}_{\rm phot} = - \frac{1}{2}  \sum_{f} &\frac{\left(m_{f, \rm tot, pred} - m_{f, \rm tot, obs}\right)^2}{\sigma^2_{f, \rm tot, obs}} \\
  &- \frac{1}{2}  \sum_{f_2} \frac{\left(\Delta m_{f_2, \rm pred} -\Delta m_{f_2, \rm obs}\right)^2}{\sigma^2_{f_2, \rm obs}}
\end{split}
\label{eqn:lnL_sed}
\end{equation} 
where the first summation is taken over the 2MASS and Pan-STARRS bands/filters ($f$ = J, H, Ks, g, r, i, z), $m_{f, \rm tot, pred}$ is the total predicted apparent magnitude of the two stars in the relevant band, $m_{f, \rm tot, obs}$ is the observed value, and $\sigma_{f, \rm tot, obs}$ is the corresponding error. Similarly,  the second term fits the flux ratios between the stars ($\Delta m = m_2 - m_1$) in the PHARO JHKs and Gaia G bands ($f_2$ = J, H, Ks, G). 

\subsection{RV model} \label{ssec:rv_model}

Finally, we use a standard Keplerian model to fit the RVs. This takes in the orbital period $P_{\rm orb}$, periastron time $t_p$, eccentricity $e$, argument of periastron $\omega$, center-of-mass RV $\gamma$, and RV semi-amplitude $K_{2}$ to predict the RV at any given time. However, since the inclination is constrained by the pulses, we directly fit for the WD mass $M_{\rm WD}$ instead of $K_{2}$. Note also that $t_p$ is a transformation of the mid-eclipse time, $t_0$, so we do not fit for it separately. The log-likelihood is defined as:
\begin{equation} \label{eqn:logL_RV}
    \ln \mathcal{L}_{\rm RV} = - \frac{1}{2}  \sum_i \frac{\left({\rm RV}_{\rm pred}\left(t_i\right) - {\rm RV}_i\right)^2}{\sigma^2_{{\rm RV}, i}}
\end{equation}
where ${\rm RV}_{\rm pred}\left(t_i\right)$ and ${\rm RV}_i$ are the predicted and measured RVs at a time $t_i$, and $\sigma_{{ \rm RV}, i}$ is the corresponding error in the measured RV. 

In summary, we fit for 14 parameters: $t_{\rm 0}$, $c_1$, $c_2$, $i$, $P_{\rm orb}$, $e$, $\omega$, $\gamma$, $M_{\rm WD}$, $M_{1}$, $M_{2}$, EEP$_1$, EEP$_2$, and $d$. The total log-likelihood function is the sum of equations \ref{eqn:logL_pulse}, \ref{eqn:lnL_sed}, and \ref{eqn:logL_RV}. 

\section{Results} \label{sec:results}

The resulting fit to the RVs, along with the residuals, are plotted in Figure \ref{fig:rv_models}. For reference, we also plot the solution from \citet{Masuda2019ApJL}. We see that neglecting the spectral contribution of the second luminous star when measuring RVs led \citet{Masuda2019ApJL} to significantly underestimate the RV semi-amplitude of the WD companion. We obtain an updated WD mass of $M_{\rm WD} = 0.53 \pm 0.01 \,M_{\odot}$, significantly higher than $0.2\, M_{\odot}$. As described in Section \ref{ssec:tres_spectra}, the RVs of star 2 used here are those obtained from fixing the RV of star 1 to a constant value. Using the RVs from fitting both stars has no effect on the best-fit WD mass. The table of best-fit values of all parameters for this fit can be found in Appendix \ref{appendix:tests}. 

\begin{figure}
    \centering
    \includegraphics[width=0.95\columnwidth]{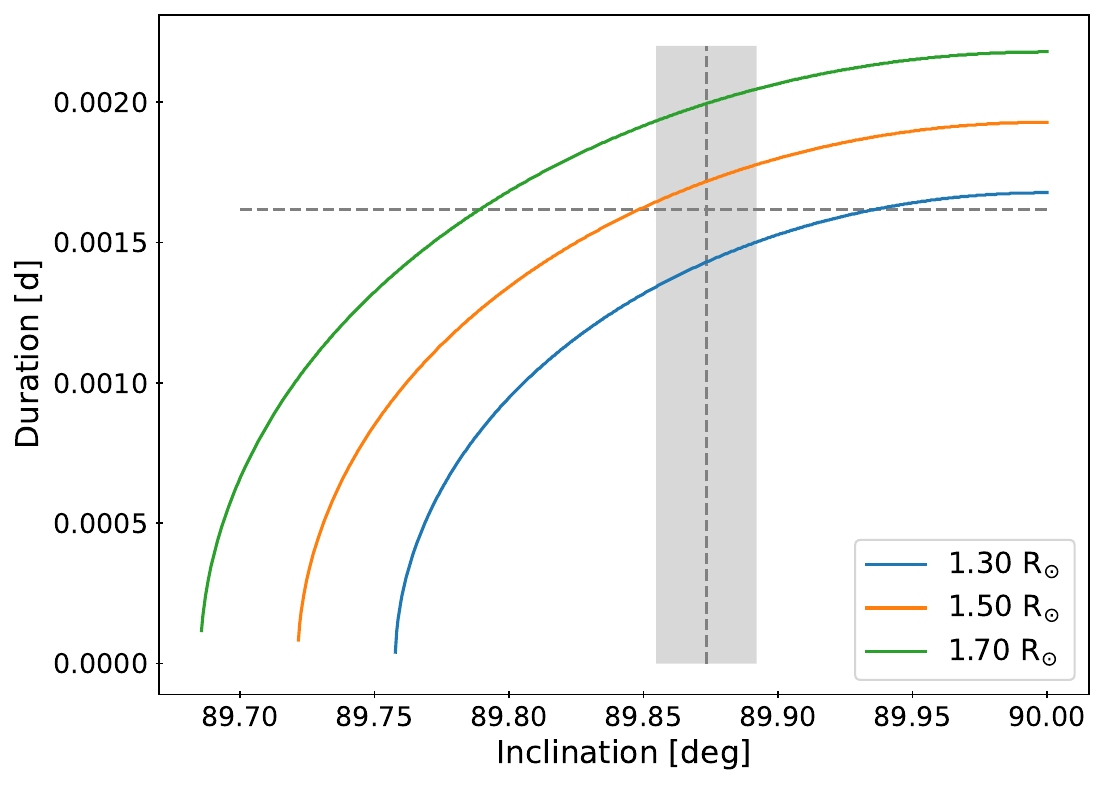}
    \caption{Dependence of the pulse duration on the inclination and the radius of the lensed star (star 2). The vertical dashed line plots the best-fit inclination, with the shaded region showing the $1-\sigma$ error. The horizontal dashed line marks the corresponding duration of the pulse. This duration cannot be matched for any inclination for $R \lesssim 1.3\,R_{\odot}$, implying that the star must be somewhat evolved.}
    \label{fig:pulse_duration}
\end{figure}

\begin{figure*}
    \centering
    \includegraphics[width=0.9\textwidth]{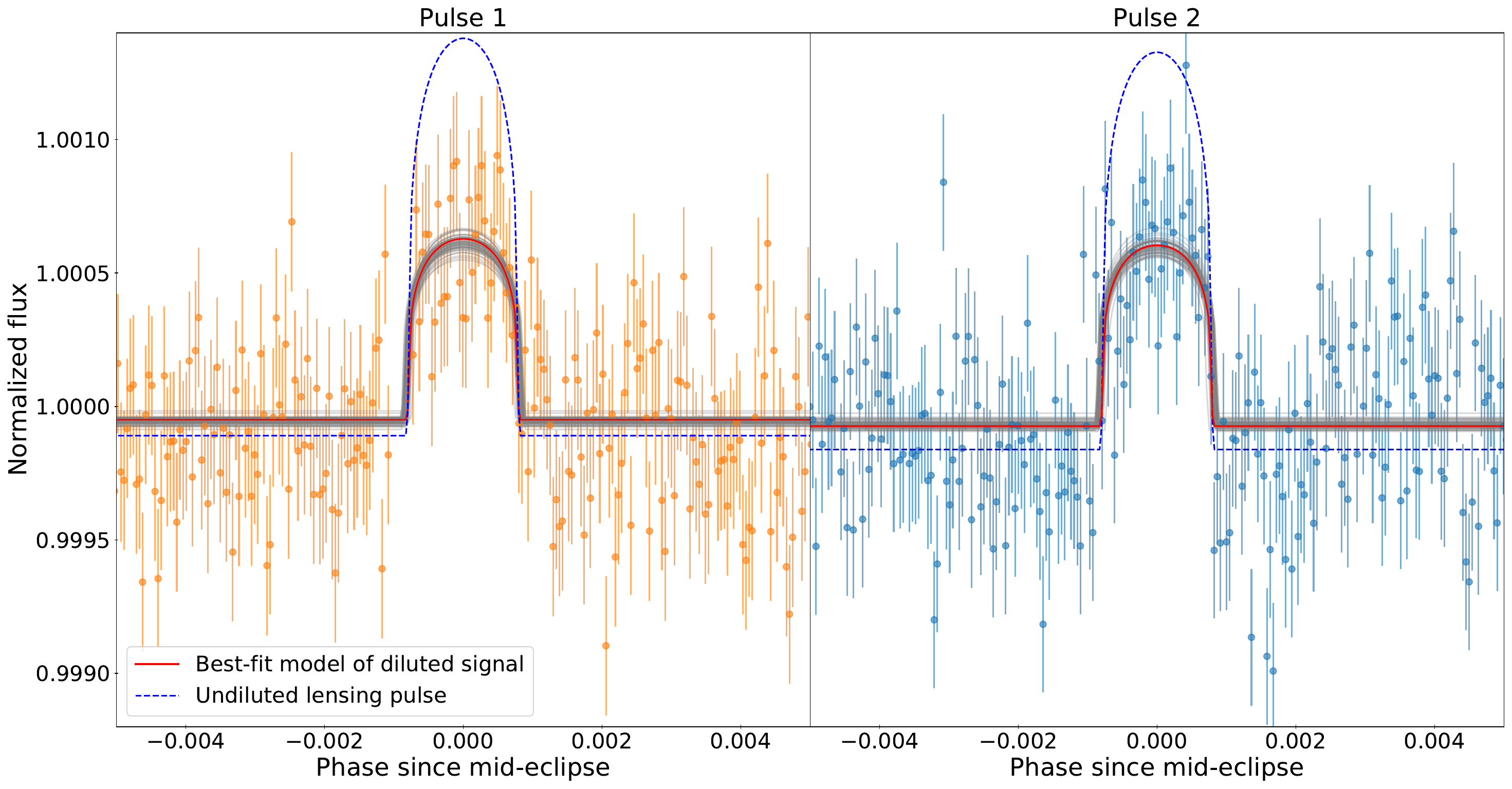}
    \caption{The two transits (de-trended) observed by Kepler around mid-eclipse, plotted against phase. The best-fit pulse model from our analysis (red solid line), and models of parameters from 50 random draws of the posteriors (gray solid lines), have been plotted on top of the data. We also plot the undiluted pulse on top (blue dashed line).}
    \label{fig:pulse}
\end{figure*}

On Figure \ref{fig:pulse}, we plot the resulting best-fit model of the pulse as well as models from 50 random draws of the posterior, along with the observed light curve for the two transits. We also plot the model pulse without dilution of light due to star 1. Since star 1 dominates the total flux, we see that the amplitude of the pulse is significantly reduced. The duration of the pulse places a strong constraint on the radius of the lensed star (star 2). The relationship between the pulse duration, inclination, and radius of the lensed star is shown in Figure \ref{fig:pulse_duration} (all other parameters were kept constant at their best-fit values).

\begin{figure}
    \centering
    \includegraphics[width=0.97\columnwidth]{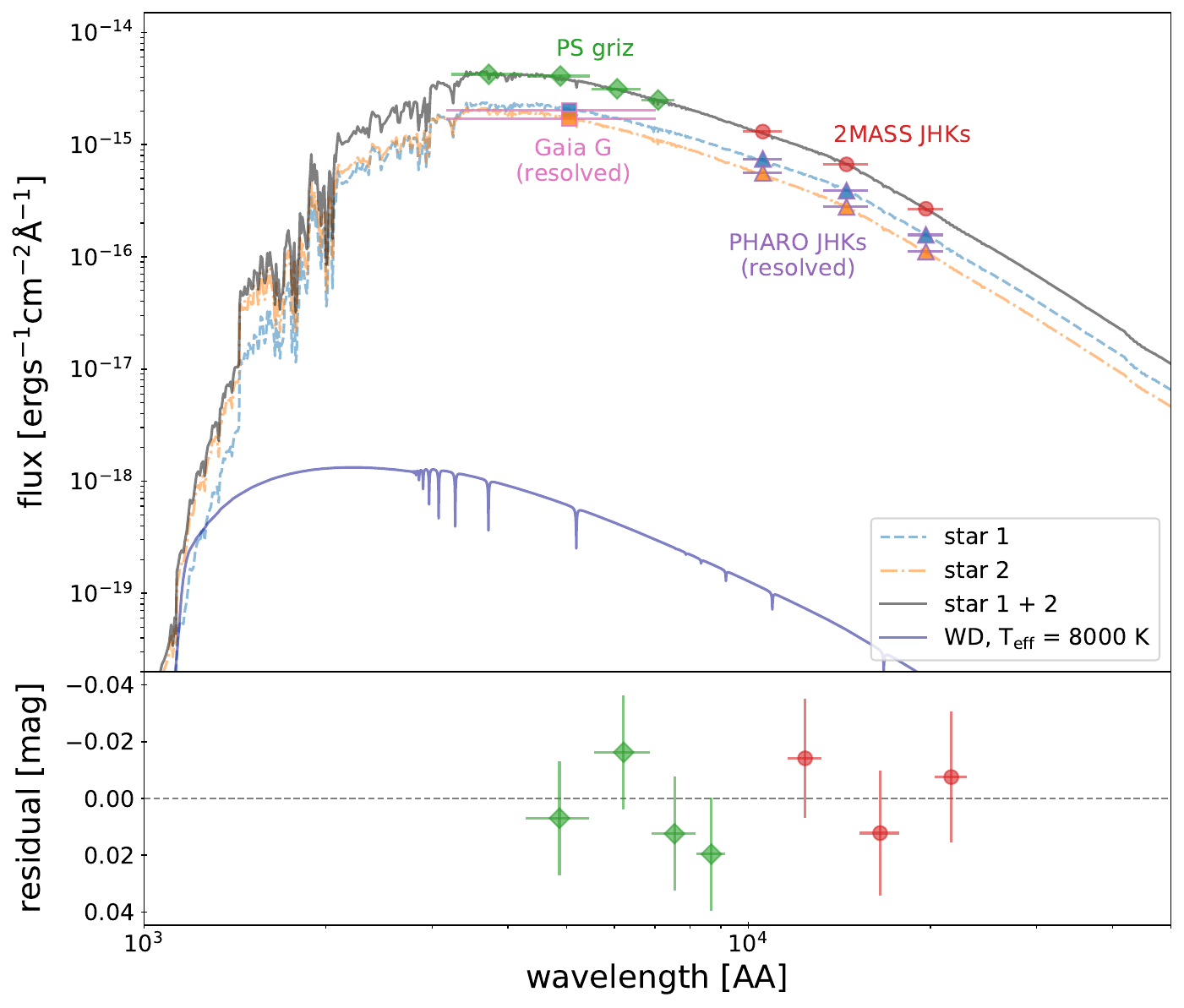}
    \caption{Model SEDs of star 1 and 2 individually, and their sum. The observed photometric points (unresolved/sum: Pan-STARRS griz, 2MASS JHKs; resolved: Gaia G, PHARO JHKs) used in the fitting have also been plotted. The error bars in wavelength represent the width of the passbands. In the lower panel, we plot the residuals (observed - model) of the unresolved points in magnitudes. A koester WD SED model has also been plotted, with an effective temperature of 8000 K, at which it contributes $\sim$0.01\% of the light in the optical.}
    \label{fig:seds}
\end{figure}

\begin{figure}
    \centering
    \includegraphics[width=0.98\columnwidth]{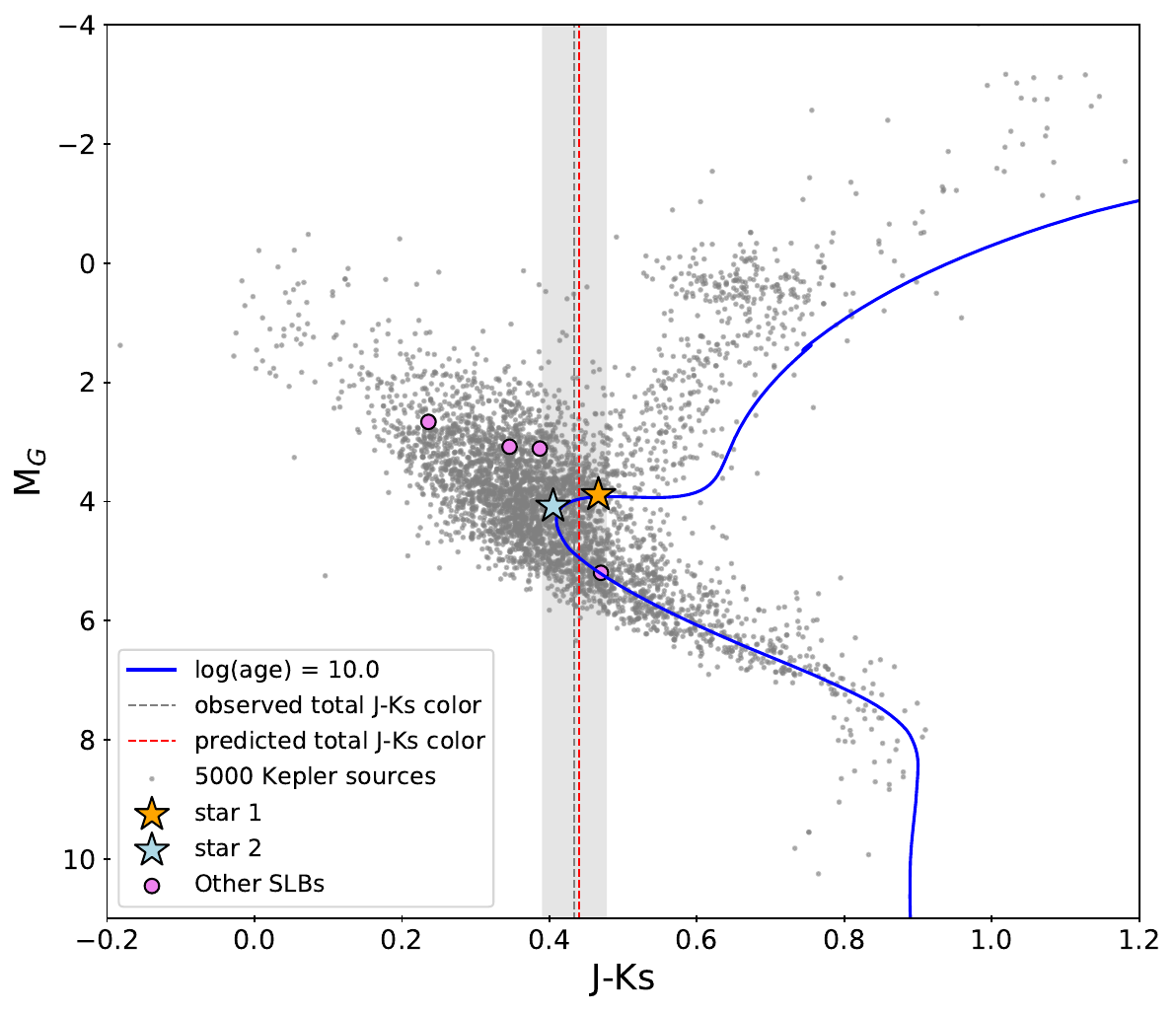}
    \caption{The locations of the two stars on a CMD of the extinction-corrected absolute Gaia G magnitude against 2MASS J-Ks color. In the background are 5000 Kepler sources and the magenta points are the four other SLBs. The MIST isochrone of star 1 is plotted with a blue solid line. The observed total J - Ks color is marked with a grey dashed line with its 1-$\sigma$ region shaded in. The predicted color is marked with a red dashed line. We see that the stars have evolved off the MS.}
    \label{fig:isochrone}
\end{figure}

Figures \ref{fig:seds} and \ref{fig:isochrone} show the results of fitting the photometry. On Figure \ref{fig:seds}, we show the model SEDs of the individual stars and their total, along with the observed Pan-STARRS, 2MASS, Gaia, and PHARO photometry. The models plotted were generated using \texttt{pytstellibs}\footnote{https://mfouesneau.github.io/pystellibs/} with the best-fit parameters as inputs. The residuals between the observed and predicted unresolved (Pan-STARRS + 2MASS) photometry are $\lesssim 0.04$ mags. For reference, we also plot a Koester WD model with $T_{\rm eff, WD} = 8000$K (and log(g) = 8.0). At this temperature, the WD contributes about 0.01\% of the light in the optical compared to the luminous stars, making it a rough upper limit based on the null detection of the secondary eclipse in the light curve. 

Figure \ref{fig:isochrone} shows the location of the two stars on a CMD of the Gaia G absolute magnitude against the 2MASS J - Ks color. Looking at the MIST isochrone (blue solid line), we see that the two stars are slightly evolved off the MS. This is also seen in comparison with the locations of 5000 random Kepler sources with apparent G magnitudes $<$ 15.5 and \texttt{parallax\_over\_error} $>$ 5, plotted in gray. The locations of the four other Kepler SLBs are plotted in magenta. As mentioned above, the location on the CMD is primarily determined by the duration of the pulse which sets the radius (Figure \ref{fig:pulse_duration}). Furthermore, the isochrones provide a constraint on the distance to the system of $1.93 \pm 0.08$ kpc. No parallax-based distance was available because both Gaia sources have 2-parameter solutions.

Finally, in Table \ref{tab:bestfit_params}, we report the best-fit values for all 14 parameters (both the values at the peaks of the probability distribution, which we refer to throughout the text, as well as the median) along with the standard deviations of the posterior (the resulting corner plot of these parameters can be found in Appendix \ref{appendix:corner}). We also summarize the key stellar parameters of the three components of the system in Table \ref{tab:stars_summary}. Note here that the effective temperatures of star 1 and star 2 are roughly consistent with those of the templates used during the cross-correlation of the spectra (Section \ref{ssec:tres_spectra}). 

\begin{table}
    \centering
    \begin{tabular}{c c c c}
         \hline
          & peak & median & stddev \\
         \hline
         $t_0$ [BJDTDB - 2454833] & 267.89 & 267.88 & 0.02 \\
         $c_1$ & 0.99989 & 0.99988 & 0.00003\\
         $c_2$ & 0.99984 & 0.99985 & 0.00003 \\
         $i$ [deg] & 89.87 & 89.88 & 0.02 \\
         $P$ [d] & 455.83 & 455.83 & 0.01 \\
         $e$ & 0.11 & 0.11 & 0.01 \\
         $\omega$ [deg] & -96.62 & -96.39 & 1.41 \\
         $\gamma$ [$\mbox{km s}^{-1}$] & -27.72 & -27.68 & 0.06 \\
         $M_{\mathrm{WD}}$ [$M_{\odot}$] & 0.53 & 0.54 & 0.01 \\
         $M_1$ [$M_{\odot}$] & 0.98 & 1.02 & 0.03 \\
         $M_2$ [$M_{\odot}$] & 0.96 & 1.00 & 0.03 \\
         EEP$_1$ & 459.75 & 459.61 & 0.64 \\
         EEP$_2$ & 447.65 & 445.93 & 1.57 \\
         $d$ [kpc] & 1.93 & 1.99 & 0.08 \\
         \hline
    \end{tabular}
    \caption{Best-fit values of the 14 parameters from the joint fitting of light curve, RVs, and photometry. We report both the values at peak of the probability distribution and the median value of the posterior. The standard deviation of the posterior is also given.}
    \label{tab:bestfit_params}
\end{table}

\begin{table*}
    \centering
    \begin{tabular}{c c c c}
         \hline
          & Star 1 & Star 2 & WD \\
         \hline
         Gaia DR3 source ID & 2105324940517591808 & 2105324936217850624 &  \\
         M [$M_{\odot}$] & 0.98 $\pm$ 0.03 & 0.96 $\pm$ 0.03 & 0.53 $\pm$ 0.01 \\
         EEP & 459.75 $\pm$ 0.64 & 447.65 $\pm$ 1.57 \\
         Age [Gyr] & 11.19 $\pm$ 1.40 & 11.58 $\pm$ 1.44 \\
         R [$R_{\odot}$] & 1.75 $\pm$ 0.06 & 1.43 $\pm$ 0.05 \\
         $T_{\rm eff}$ [K] & 5393 $\pm$ 80 & 5683 $\pm$ 86 & $\lesssim 8000$ \\
         \hline
    \end{tabular}
    \caption{Summary of the stellar parameters for the three stars in KIC 8145411. The age, radii and effective temperatures of the two luminous components were obtained from isochrones given the best-fit parameters from the joint fitting (Section \ref{sec:fitting}). The errors are the standard deviations of the values obtaining by taking parameters from 50 random draws of the posteriors. The WD temperature is a rough upper limit based on the SED (Figure \ref{fig:seds}).}
    \label{tab:stars_summary}
\end{table*}

\subsection{Formation history}

\begin{figure*}
    \centering
    \includegraphics[width=0.75\textwidth]{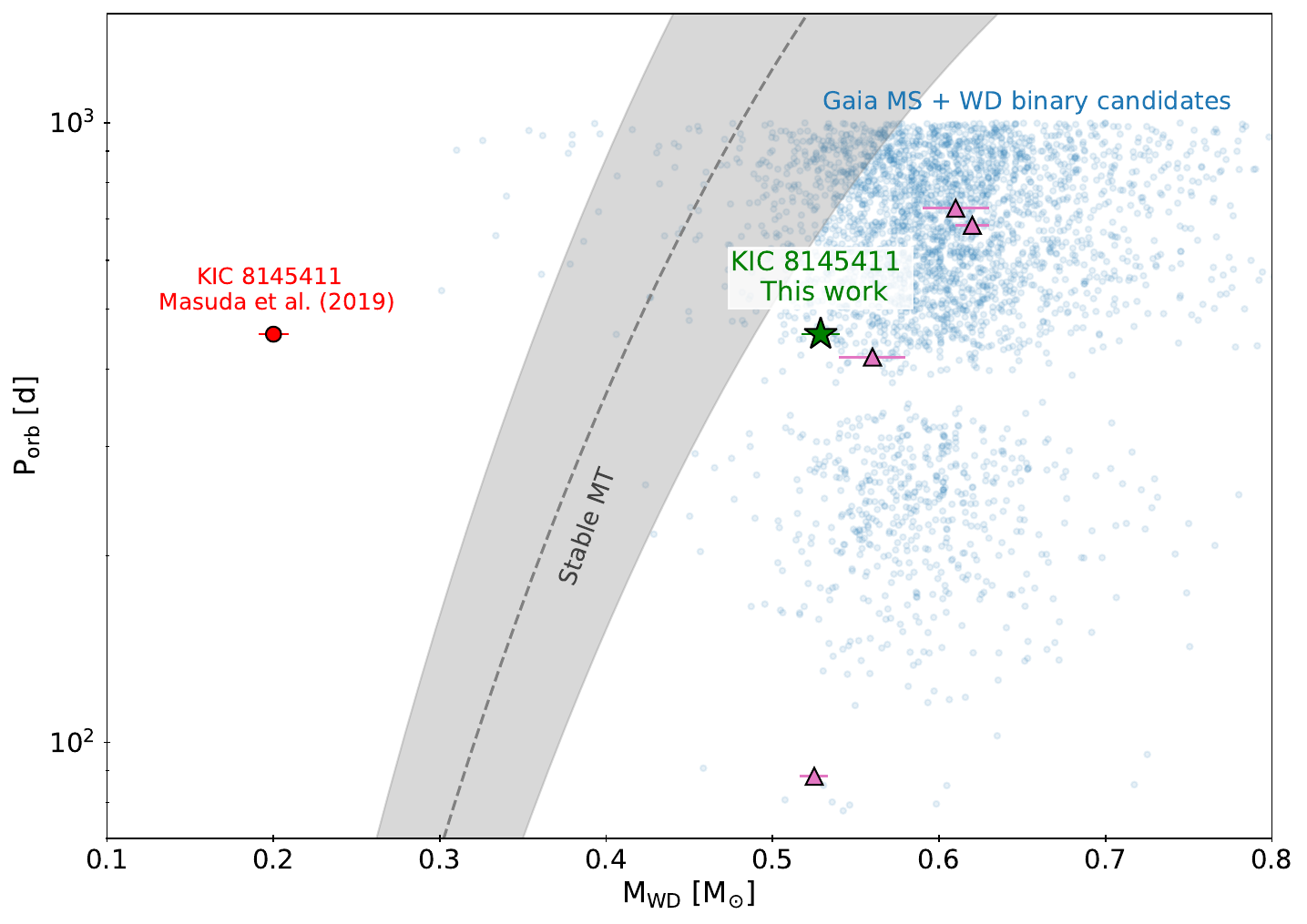}
    \caption{Orbital period against WD mass for MS + WD binaries from the literature. The location of KIC 8145411 from \citet{Masuda2019ApJL} is shown with a red circle marker, while the new solution from this work has been plotted with a green star. The four other SLBs -- KOI-3278 from \citealt{Kruse2014Sci}, and the other three from \citealt{Kawahara2018AJ} -- are shown with triangle markers (with updated parameters from \citealt{Masuda2020IAUS} and \citealt{Yahalomi2019ApJ}). We also plot the sample of MS + WD binary candidates identified by \citet{Shahaf2024MNRAS} from Gaia DR3. Binaries formed from stable MT are expected to evolve along the dashed line \citep{Rappaport1995MNRAS}. Known PCEBs are located further below this plot at orbital periods $\lesssim 1$d \citep[e.g.][]{Rebassa-Mansergas2007MNRAS}, while the ``wide" PCEBs \citep{Wonnacott1993, Yamaguchi2024MNRAS} hosting ultramassive WDs are located further to the right.}
    \label{fig:P_v_Mwd}
\end{figure*}

Through our analysis, we showed that KIC 8145411 is not a single binary hosting an ELM WD, but that it is a triple, with an inner binary containing a solar-type star and a $0.53\,M_{\odot}$ WD, and a luminous tertiary that is another solar-type star. As shown on the $P_{\rm orb}-M_{\rm WD}$ plot of Figure \ref{fig:P_v_Mwd}, this moves the system from lying above the stable MT line on the left to the right, much closer to the other SLBs and many of the MS + WD binary candidates from the Gaia astrometric sample \citep{Shahaf2024MNRAS}. This resolves the mystery of the ``impossible" ELM WD as it no longer needs to have undergone significant mass loss through stable MT when the progenitor was an early RGB star. 

However, like many of the other objects shown on Figure \ref{fig:P_v_Mwd}, the system now lies below the standard $P_{\rm orb}-M_{\rm WD}$ relation that stable MT binaries are expected to evolve along, plotted as a gray dashed line in Figure \ref{fig:P_v_Mwd} (the shaded region represents an uncertainty of a factor of 2.4; \citealt{Rappaport1995MNRAS}). Similar to the other SLBs and Gaia MS + WD binary candidates \citep{Shahaf2024MNRAS}, some uncertainty remains in the formation history of KIC 8145411. It must have undergone interaction prior to the formation of the WD but its orbit is smaller than expected for a post-stable MT system. Meanwhile, its orbit is larger than PCEBs predicted by previous binary population synthesis codes \citep[e.g.][]{Zorotovic2014A&A}, though it is consistent with some recent models of CEE from thermally-pulsating AGB donors (e.g. \citealt{Yamaguchi2024arXiv, Yamaguchi2024MNRAS}; \citealt{Belloni2024arXiv, Belloni2024arXiv_b}). 

The SLBs are a great complementary sample to Gaia as they probe a similar region of the $P_{\rm orb}-M_{\rm WD}$ space, while having a simpler selection function that depend primarily on the eclipse probability. As described in \citet{Masuda2019ApJL}, given that 5 SLBs have so far been found, we can roughly estimate that $\sim 1 \%$ of all Sun-like stars host WDs in AU-scale orbits. Given a local stellar density of $\sim 0.1 \,\mbox{pc}^{-3}$ \citep[e.g.][]{Golovin2023A&A} and local WD density of $4.47 \times 10^{-3}\, \mbox{pc}^{-3}$ \citep{GentileFusillo2021MNRAS}, this translates to $\sim 20 \%$ of all WDs being found in these binaries. 

\section{Conclusions} \label{sec:conclusion}

In this work, we showed that the self-lensing binary KIC 8145411 is an inner binary of a hierarchical triple. Neglecting the light from the tertiary both reduces the measured RV semi-amplitude of the lensed star, and dilutes the self-lensing pulse in the light curve, which both act to reduce the inferred WD mass. Therefore, the originally reported WD mass of $\sim 0.2 M_{\odot}$ by \citet{Masuda2019ApJL} was underestimated and we obtain an updated mass of $0.53 \pm 0.01 M_{\odot}$.

This resolves the challenge of forming an ELM WD in a wide orbit ($P_{\rm orb} = 455.83 \pm 0.01$). However, the system now has a period and WD mass comparable to the other SLBs, as well as the large sample of MS + WD binary candidates from \citet{Shahaf2024MNRAS}, whose mass transfer histories remain unknown. 

We note that since a significant fraction of solar-type stars have wide binary companions at distances $\gtrsim 100\,$AU \citep[e.g.][]{Rastegaev2010AJ, El-Badry2021MNRAS}, the problem of neglected contamination from an unresolved outer star is likely to be common. In fact, the same effect has been identified to lead to underestimated planet radii from transits \citep{Ciardi2015ApJ}. High-resolution imaging is an effective tool to mitigating this problem.

\section{Acknowledgments}

NY and KE acknowledge support from NSF grant AST-2307232. 

This research was carried out in part at the Jet Propulsion Laboratory, California Institute of Technology, under a contract with the National Aeronautics and Space Administration (80NM0018D0004).

This work has made use of data from the European Space Agency (ESA) mission
{\it Gaia} (\url{https://www.cosmos.esa.int/gaia}), processed by the {\it Gaia}
Data Processing and Analysis Consortium (DPAC,
\url{https://www.cosmos.esa.int/web/gaia/dpac/consortium}). Funding for the DPAC
has been provided by national institutions, in particular the institutions
participating in the {\it Gaia} Multilateral Agreement.

This paper includes data collected by the Kepler mission and obtained from the MAST data archive at the Space Telescope Science Institute (STScI). Funding for the Kepler mission is provided by the NASA Science Mission Directorate. STScI is operated by the Association of Universities for Research in Astronomy, Inc., under NASA contract NAS 5–26555.

The Pan-STARRS1 Surveys (PS1) and the PS1 public science archive have been made possible through contributions by the Institute for Astronomy, the University of Hawaii, the Pan-STARRS Project Office, the Max-Planck Society and its participating institutes, the Max Planck Institute for Astronomy, Heidelberg and the Max Planck Institute for Extraterrestrial Physics, Garching, The Johns Hopkins University, Durham University, the University of Edinburgh, the Queen's University Belfast, the Harvard-Smithsonian Center for Astrophysics, the Las Cumbres Observatory Global Telescope Network Incorporated, the National Central University of Taiwan, the Space Telescope Science Institute, the National Aeronautics and Space Administration under Grant No. NNX08AR22G issued through the Planetary Science Division of the NASA Science Mission Directorate, the National Science Foundation Grant No. AST–1238877, the University of Maryland, Eotvos Lorand University (ELTE), the Los Alamos National Laboratory, and the Gordon and Betty Moore Foundation.

\vspace{5mm}
\facilities{FLWO:1.5m, Hale, Gaia, Kepler}

\software{pytransit \citep{Parviainen2015MNRAS}, emcee \citep{Foreman-Mackey2013PASP}}

\appendix

\section{PHARO imaging supplementary material} \label{appendix:pharo_extra}

Table \ref{tab:sep_PA} lists the angular separations between the two luminous components (star 1 and 2) and their position angles from our PHARO observations in the J, H, and Ks bands. 

\begin{table} [htb!]
    \centering
    \begin{tabular}{c c c c}
         \hline
         & & Separation [arcsec] & PA [deg] \\
         \hline
         PHARO & J & 0.321 $\pm$ 0.002 & 220.26 $\pm$ 0.63 \\
         & H & 0.323 $\pm$ 0.002 & 219.98 $\pm$ 0.63 \\
         & Ks & 0.323 $\pm$ 0.002  & 219.98 $\pm$ 0.63 \\
         Gaia & G & 0.319  & 218.01 \\
         \hline
    \end{tabular}
    \caption{Angular separations between the luminous components and their position angles based on the coordinates measured by our PHARO observations in the J, H, and Ks bands. We also quote the values calculated using the Gaia coordinates of the sources.}
    \label{tab:sep_PA}
\end{table}

\begin{figure*} [htb!]
    \centering
    \includegraphics[width=0.7\textwidth]{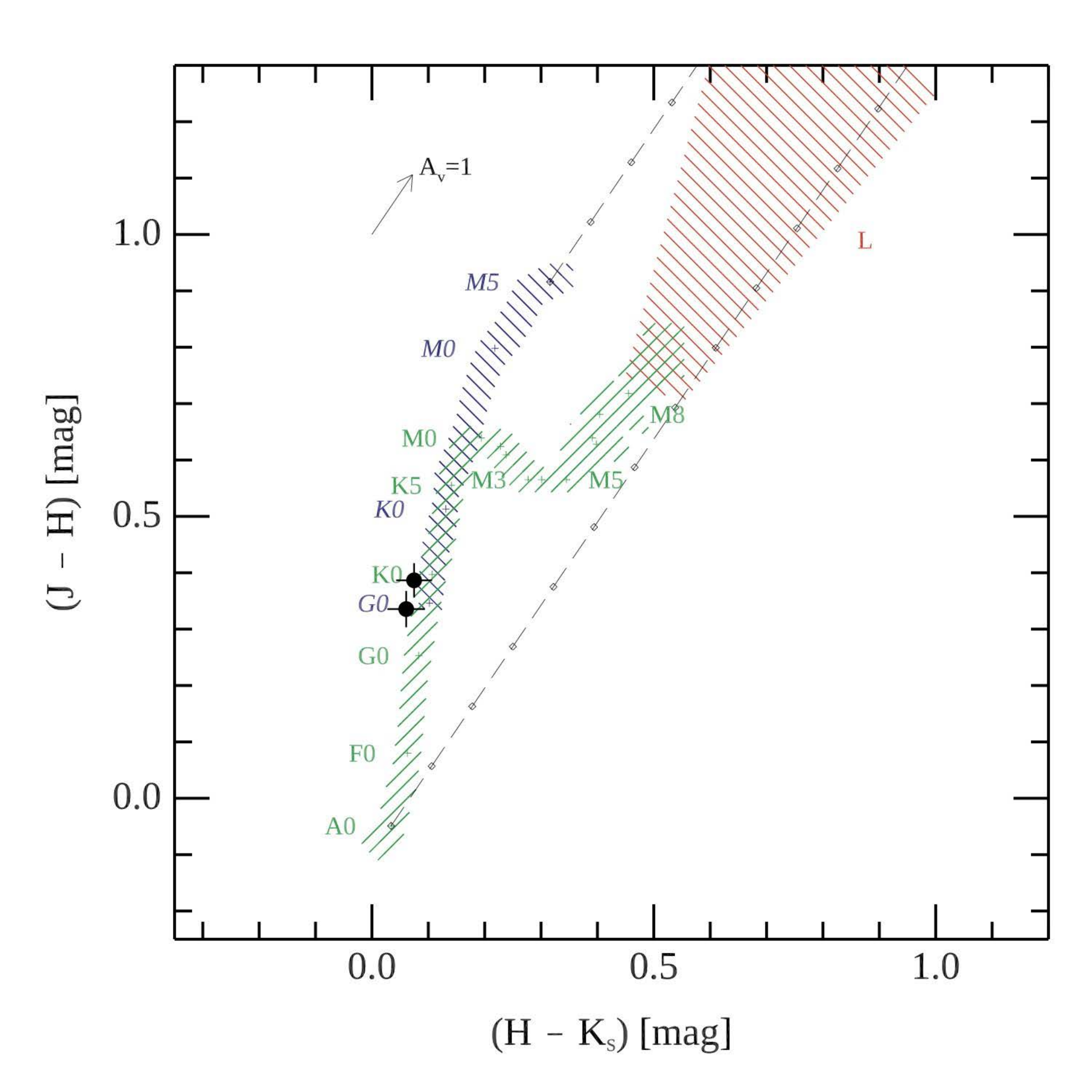}
    \caption{The locations of the two luminous stars (black markers) on a JHKs color-color plot from the PHARO observation (Section \ref{ssec:PHARO}). The dashed regions show the intrinsic colors for MS stars (blue), giants (green), and supergiants (red). The reddening vector for $A_V = 1$ is shown with a grey arrow.}
    \label{fig:jhks_plot}
\end{figure*}

Figure \ref{fig:jhks_plot} shows the locations of the two stars as observed by PHARO (Section \ref{ssec:PHARO}) on a JHKs color-color plot. The blue and green dashed regions are the intrinsic (unreddened) colors for MS stars and giants, respectively. We see that the resolved photometry is roughly consistent with the results of the isochrone fitting (Section \ref{sec:results}). 

\section{Radial velocities across orders} \label{appendix:rvs_across_orders}

\begin{figure*}
    \centering
    \includegraphics[width=0.48\textwidth]{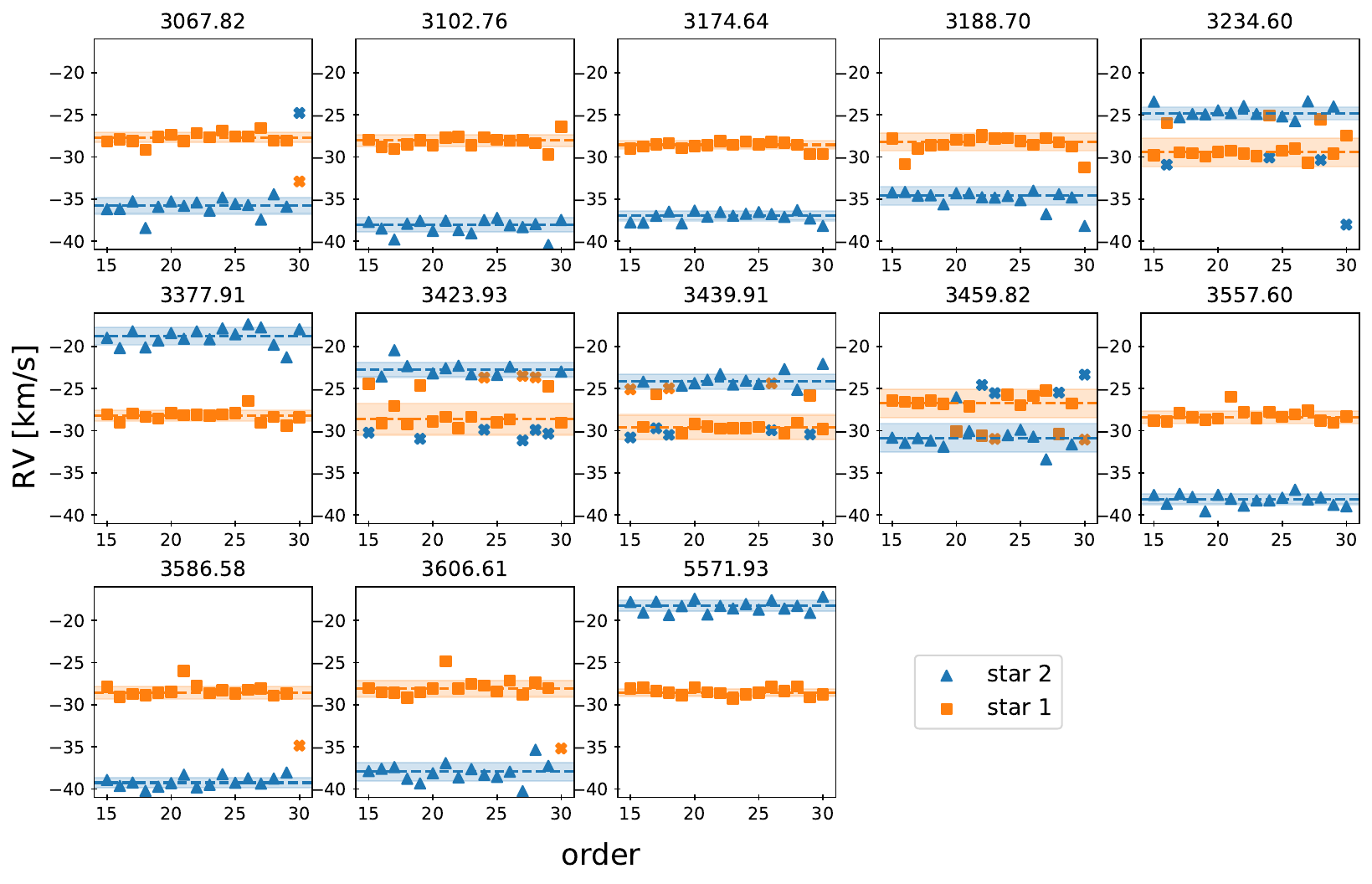}
    \includegraphics[width=0.48\textwidth]{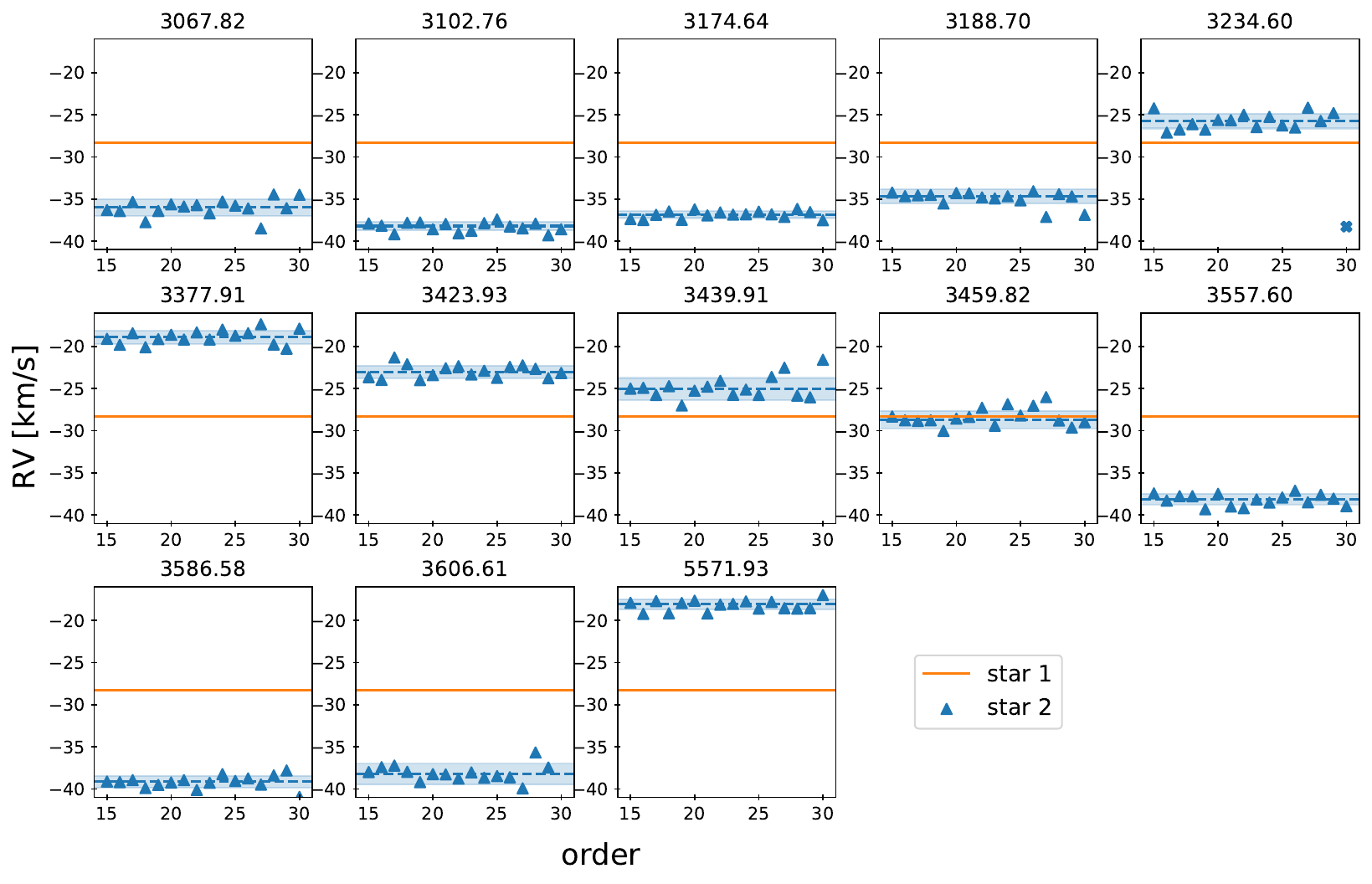}
    \caption{Measured RVs of star 1 (orange squares) and star 2 (blue triangles) against the orders, across epochs (BJD - 2454833). Points that are marked with crosses are removed if the outlier cut is implemented. The median values are plotted with dashed lines, and the shaded region corresponds to  the 1-$\sigma$ spread in the RVs (the errors in the median are taken to be this divided by the number of orders). On the right, the RVs of both stars are varied and fitted. We see that the RVs of star 1 remain roughly constant across epochs, indicating that it is the outermost star. At epochs where the RV curves of the two stars cross, we see that there is degeneracy between a given pair of RVs and the reversed pair. On the left, the RV of star 1 has been fixed at $-28.3 \mbox{km s}^{-1}$, and just the RVs of star 2 are fitted.}
    \label{fig:rvs_across_orders}
\end{figure*}

On Figure \ref{fig:rvs_across_orders}, we plot the RVs across orders at each epoch. On the right panel, we let both the RVs of star 1 and star 2 vary, while on the left panel, we fix the RV of star 1 to a constant value. Points that deviate by more than 15\% from the median are taken to be outliers and removed. These are identified with cross markers on the plots.   

\section{Best-fit parameters for different tests} \label{appendix:tests}

Tables \ref{tab:bestfit_params_tests} and \ref{tab:bestfit_params_tests_2} report the best-fit values of all 14 parameters from the joint fitting of the light curve, RVs, and photometry for four different tests: using RVs from fitting the RVs of both stars with and without outlier removal (as opposed to fixing those of star 1; Section \ref{ssec:tres_spectra}), using limb-darkening coefficients for a 5500 K star (Section \ref{ssec:pulse_model}), and removing the age constraint on the two stars ($|$log(age$_1$) - log(age$_2$)$|$ $< 0.02$; Section \ref{ssec:sed_model}). We see that the best-fit value for the WD mass is robust to all of these assumptions. 

\begin{table} 
    \centering
    \begin{tabular}{c | c c c | c c c}
        \hline 
         & \multicolumn{3}{c|}{With outlier cut} & \multicolumn{3}{c}{Without outlier cut} \\
          & peak & median & stddev  & peak & median & stddev  \\
         \hline
         $t_0$ [BJDTDB - 2454833] & 267.89 & 267.88 & 0.02 & 267.88 & 267.88 & 0.02  \\
         $c_1$ & 0.99989 & 0.99988 & 0.00003 & 0.99988 &	0.99988 &	0.00003 \\
         $c_2$ & 0.99986 & 0.99985 & 0.00003 & 0.99985 &	0.99985 &	0.00003 \\
         $i$ [deg] & 89.87 & 89.88 & 0.02 & 89.88 & 89.89 & 0.02 \\
         $P$ [d] & 455.83 & 455.83 & 0.01 & 455.83 & 455.83 & 0.01 \\
         $e$ & 0.12 & 0.12 & 0.01 & 	0.10 & 0.11 & 0.01  \\
         $\omega$ [deg] & -100.09 & -100.12 & 1.40 & -90.85 & -91.56 & 2.85  \\
         $\gamma$ [$\mbox{km s}^{-1}$] & -27.60 & -27.57 & 0.06 & -27.83 & -27.83 & 0.08 \\
         $M_{\mathrm{WD}}$ [$M_{\odot}$] & 0.53 & 0.54 & 0.01 & 0.52 & 0.53 & 0.01  \\
         $M_1$ [$M_{\odot}$] & 0.98 & 1.02 & 0.03 & 0.98 & 1.00 & 0.03 \\
         $M_2$ [$M_{\odot}$] & 0.95 & 1.00  & 0.03 & 0.96 & 0.98 & 0.03\\
         EEP$_1$ & 459.91 & 459.57 & 0.62 & 459.82 &	459.55 & 0.63 \\
         EEP$_2$ & 448.06 & 445.90 & 1.55 & 447.45 &	445.92 & 1.66 \\
         $d$ [kpc] & 1.93 & 1.99 & 0.07 & 1.93 & 1.97 & 0.07 \\
         \hline
    \end{tabular}
    \caption{Same as Table \ref{tab:bestfit_params}, but showing results from the joint fitting using RVs of star 2 obtained from fitting the RVs of both stars in the cross-correlation of the spectra (left panel of Figure \ref{fig:rv_models}; see Section \ref{ssec:tres_spectra}), with and without outlier removal.}
    \label{tab:bestfit_params_tests}
\end{table}

\begin{table} 
    \centering
    \begin{tabular}{c | c c c | c c c}
        \hline 
          & \multicolumn{3}{c|}{Different LDCs} & \multicolumn{3}{c}{No age constraint} \\
          & peak & median & stddev  & peak & median & stddev \\
         \hline
         $t_0$ [BJDTDB - 2454833] & 267.88 & 267.88 & 0.02  & 267.88 & 267.88 & 0.02 \\
         $c_1$  & 0.99987 & 0.99988 & 0.00003 & 0.99988 & 0.99988 & 0.00003 \\
         $c_2$ & 0.99985 & 0.99985 & 0.00003 & 0.99984 & 0.99985 & 0.00003 \\
         $i$ [deg] & 89.88 & 89.88 & 0.02 & 89.88 & 89.88 & 0.02 \\
         $P$ [d]  & 455.83 & 455.83 & 0.01 & 455.83 & 455.83 & 0.01 \\
         $e$ & 0.11 & 0.12 & 0.01 & 0.11 & 0.12 & 0.01 \\
         $\omega$ [deg] & -99.92 & -100.11 & 1.38 & -99.81 & -100.11 & 1.44 \\
         $\gamma$ [$\mbox{km s}^{-1}$]  & -27.56 & -27.57 & 0.07 & -27.57 & -27.57 & 0.06 \\
         $M_{\mathrm{WD}}$ [$M_{\odot}$] & 0.54 & 0.54 & 0.01 & 0.53 & 0.54 & 0.01 \\
         $M_1$ [$M_{\odot}$] & 1.00 & 1.02 & 0.03 & 0.98 & 1.02 & 0.03 \\
         $M_2$ [$M_{\odot}$] & 0.98 & 1.00 & 0.03 & 0.96 & 1.00 & 0.03 \\
         EEP$_1$ & 459.21 & 459.60 & 0.65 & 459.32 & 459.71 & 0.72 \\
         EEP$_2$ & 445.33 & 445.81 & 1.50 & 446.67 & 446.50 & 2.08 \\
         $d$ [kpc] & 1.96 & 2.00 & 0.08  & 1.93 & 2.00 & 0.08 \\
         \hline
    \end{tabular}
    \caption{Same as Table \ref{tab:bestfit_params_tests}, but showing results from the joint fitting (1) using limb-darkening coefficients for a star of $T_{\rm eff} = 5500\,$K, as opposed to $5750\,$K (Section \ref{ssec:pulse_model}), and (2) without imposing that the two stars be close in age in fitting the photometry (Section \ref{ssec:sed_model}).}
    \label{tab:bestfit_params_tests_2}
\end{table}

\section{Corner plot from joint fitting} \label{appendix:corner}

\begin{figure*}
    \centering
    \includegraphics[width=0.95\textwidth]{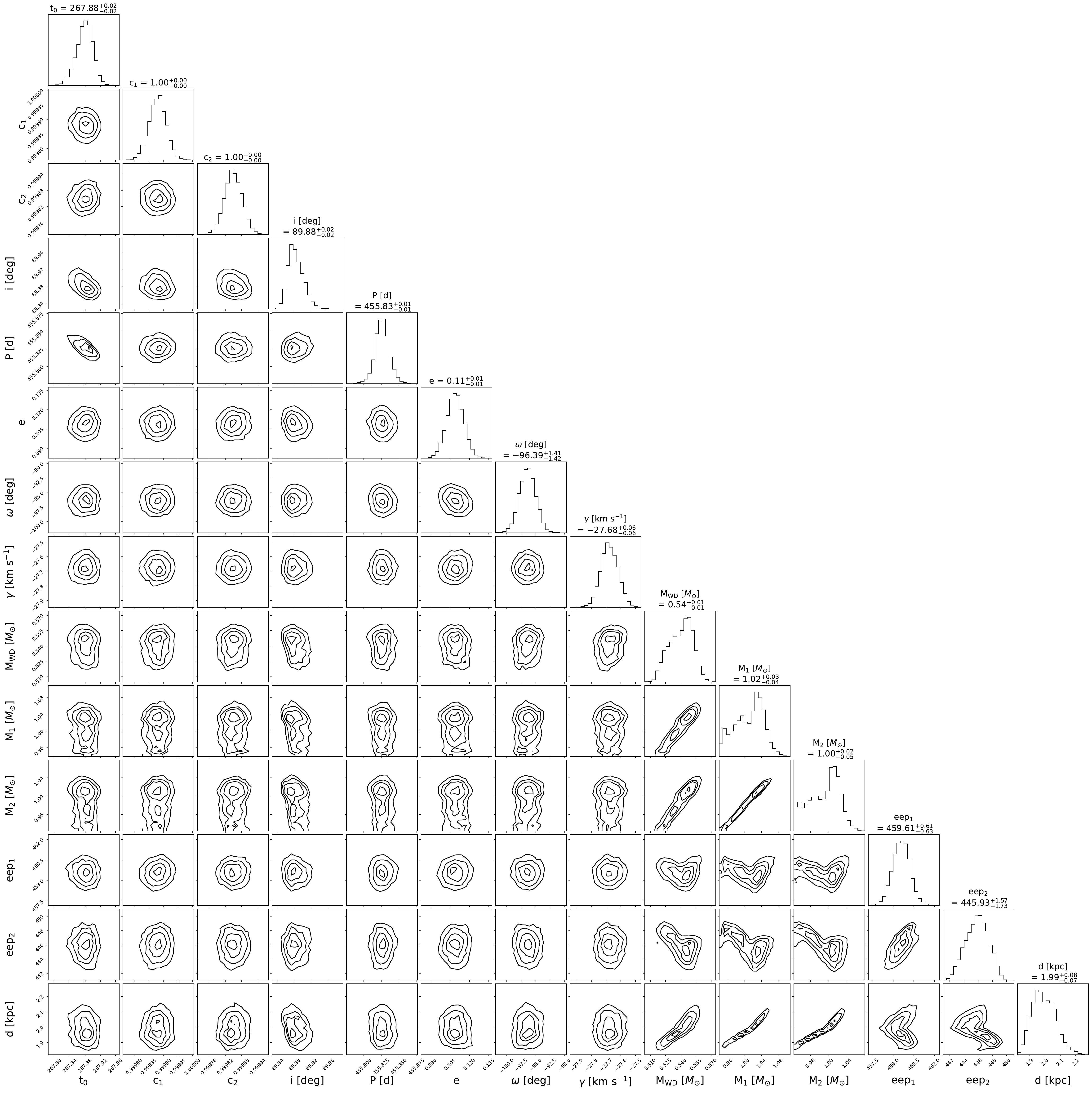}
    \caption{Corner plot showing the posterior distributions of the 14 parameters.}
    \label{fig:corner}
\end{figure*}

Figure \ref{fig:corner} is a corner plot showing the posterior distributions of 14 paramerers from the joint fitting of light curves, RVs, and photometry as described in Section \ref{sec:fitting} and whose results are discussed in Section \ref{sec:results}.  

\bibliographystyle{aasjournal}

\end{document}